\documentclass[conference]{IEEEtran}
\usepackage{cite}

\ifCLASSINFOpdf
   \usepackage[pdftex]{graphicx}
   \DeclareGraphicsExtensions{.pdf,.jpeg,.png}
\else
\fi
\usepackage{amsmath}
\usepackage{array}

  \usepackage[caption=false,font=footnotesize]{subfig}

\usepackage{stfloats}
\usepackage{url}

\usepackage{mathtools,amssymb,lipsum}

\usepackage{cuted}
\usepackage{breqn}
\allowdisplaybreaks

\newcounter{MYtempeqncnt}
\begin{document}
%
\title{Small Signal Audiosusceptibility Model for\\Series Resonant Converter}
 \author{\IEEEauthorblockN{Subhash Joshi T.G.}
 \IEEEauthorblockA{Member, IEEE\\Power Electronics Group\\
 Centre for Development of Advanced Computing\\
 Thiruvananthapuram-695033, India\\
 Email:subhashj@cdac.in}
 \and
 \IEEEauthorblockN{Vinod John}
 \IEEEauthorblockA{Senior Member, IEEE\\Department of Electrical Engineering\\
 Indian Institute of Science\\
 Bangalore-560012, India\\
 Email:vjohn@iisc.ac.in}
 }

%

\IEEEoverridecommandlockouts

\maketitle

\begin{abstract}
Models that accurately predict the output voltage ripple magnitude are essential for applications with stringent performance target for it. Impact of dc input ripple on the output ripple for a Series Resonant Converter (SRC) using discrete domain exact discretization modelling method is analysed in this paper. A novel discrete state space model along with a small signal model for SRC considering 3 state variables is presented. The audiosusceptibility (AS) transfer function which relates the input to output ripple is derived from the small signal model. Analysis of the AS transfer function indicates a resonance peak and an expression is derived connecting the AS resonance frequency for input ripple with different SRC component values. Further analysis is done to show that a set of values for SRC parameter exists, which forms a design region, for which the normalized gain offered by the SRC for input ripple is less than unity at any input ripple frequency. A test setup to introduce the variable frequency ripple at the input of SRC for the experimental evaluation of AS transfer function is also proposed. Influence of stray parameters on AS gain, AS resonance frequency and on SRC tank resonance frequency is addressed. An SRC is designed at a power level of $10kW$. The analysis using the derived model, simulations, and experimental results are found to be closely matching.
\end{abstract}
\begin{IEEEkeywords}
series resonant converter, sampled data modelling, audiosusceptibility, small signal model
\end{IEEEkeywords}

%
\IEEEpeerreviewmaketitle

\section{Introduction}
\begin{figure}[!t]
\centering
\includegraphics[keepaspectratio,scale=0.65]{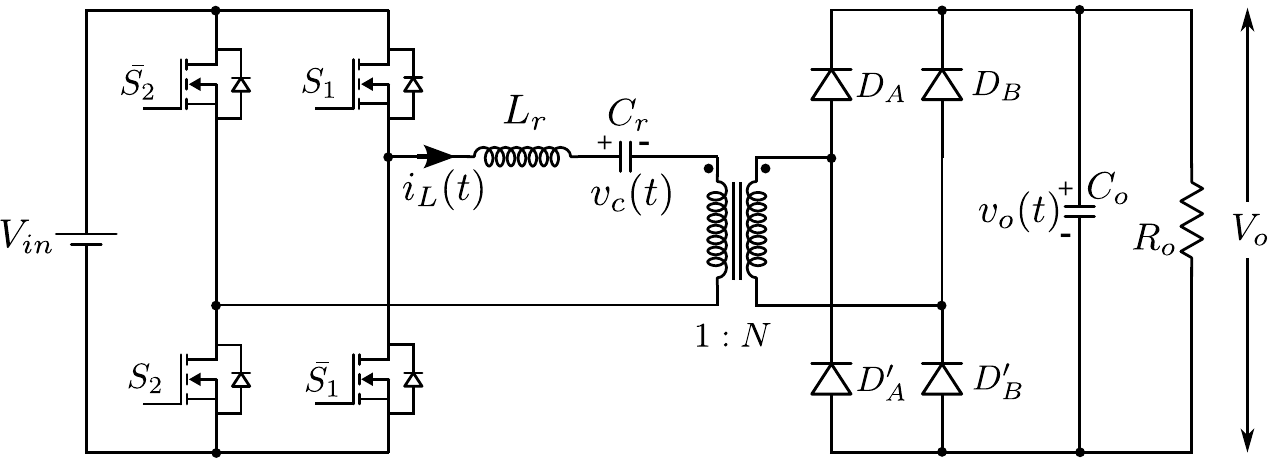}
\caption{SRC showing the state variables.}
\label{Fig1a}
\end{figure}

Series Resonant Converter (SRC), shown in Fig. \ref{Fig1a}, is the preferred topology for High Voltage (HV) low current power supply due to the absence of magnetics at the HV side~\cite{ref3}. Radar, X-rays are some of the applications where such HV power supplies are used. Some of these applications impose stringent performance parameters on the HV power supply such as voltage ripple and regulation~\cite{ref7}. To maintain better imaging quality for radar and X-ray image contrast, these applications demand a good control over dc voltage ripple of less than 0.001\%. In this paper an analysis is carried out to evaluate the contribution of input voltage ripple on output voltage ripple, which is also referred to as audiosusceptibility (AS) model~\cite{ref2}.

Small signal models of resonant converters are of interest over a long time and recent works have further refined the converter model~\cite{ref6,ref8,ref9,ref10,ref11,ref12,ref13,ref14}. In many small signal models, transfer functions are derived based on numerical solutions instead of analytical solutions that result in loss of physical insight, which hampers its use for design purpose. The widely used state-space averaging approach fails for resonant converter since the states are of ac behaviour~\cite{ref6}. Small signal models for SRC are derived by using High-Q approximation~\cite{ref8}. Another modelling approach is by the low frequency approximation of the states~\cite{ref9}. Both these approaches have sinusoidal waveform assumption that results in accuracy limitations. The small signal transfer functions of SRC in~\cite{ref10} are derived numerically due to the difficulty in converting it into an analytical expression form. In~\cite{ref11} the amplitude and the phase of the states are transformed into slowly varying signals where the accuracy is improved by finding a series of averaged models which necessitate the use of numerical approach. In extended describing function method each non-linear elements are approximated by its first harmonic sinusoids~\cite{ref12,ref13}. In contrast, the discrete domain modelling method using exact discretization, that is also referred as sampled data modelling method, can be employed without any approximation and hence it is capable of analysing the circuit accurately~\cite{ref14}. In this method the modelling is carried out by solving time segmented piece-wise linear time invariant state equations by use of the switching boundary conditions. Then by performing the perturbation and linearisation, a small signal model is accurately obtained. Due to the analysis complexity, the converter models using exact discretization modelling method reported so far only provide numerical solutions~\cite{Juergen}.

In~\cite{ref15_}, the state space model of the SRC over a switching period is obtained analytically by using exact discretization modelling approach. A simple method to handle the resonant tank and output filter state variables of the SRC while formulating the combined state space equation for the switching period $T_s$ has been proposed. From the state space model of SRC, an analytical expression for small signal model connecting the input and states is derived. The audiosusceptibility (AS) transfer function is also derived from the small signal model. The resulting model provides analytical parameter relationships and physical insights into the system in terms of SRC circuit parameters.

In this paper it is shown that it is possible to have a resonance peak in the AS transfer function of the SRC. An analytical expression is derived for the AS resonance frequency for input ripple and its relationship with different SRC components is shown. It is also demonstrated that the AS resonance frequency is much lower than the SRC tank resonance frequency which is decided by the resonant tank components. From the derived transfer function, the paper discusses the gain and open-loop bandwidth offered by the SRC to variations in input frequency. The frequency response to input ripple is verified by simulations and compared with the derived analytical expressions for different values of effective quality factor, $Q_e$, and ratio of switching frequency to SRC tank resonance frequency, $F$, where $Q_e$ is as defined in~\cite{text1}. The paper extends the analysis to show that a set of value for $F$ and $Q_e$ exists for which the normalized gain offered by SRC for input ripple is less than unity for any input ripple frequency. The paper utilizes this finding to identify a design and operating region that relate $Q_e$ and $F$ with the ``less than unity gain'' in audiosusceptibility. This region derived from the analytical AS transfer function is verified by time domain simulation results. An SRC is designed at a power level of $10kW$. An operating condition is chosen to demonstrate the AS resonance for input ripple. The paper also proposes a test setup to introduce variable frequency ripple at the input of the SRC for the experimental evaluation of AS transfer function. The influence of stray parasitic parameters on AS gain, AS resonance frequency, and on SRC tank resonance frequency are verified by the model, simulation and experiment. It is shown that the influence of stray parameter on AS resonance frequency is minimal. The analysis of the derived model, simulation and experimental results are found to be closely matching.
\begin{figure}[!b]
\centering
\includegraphics[keepaspectratio,scale=0.87]{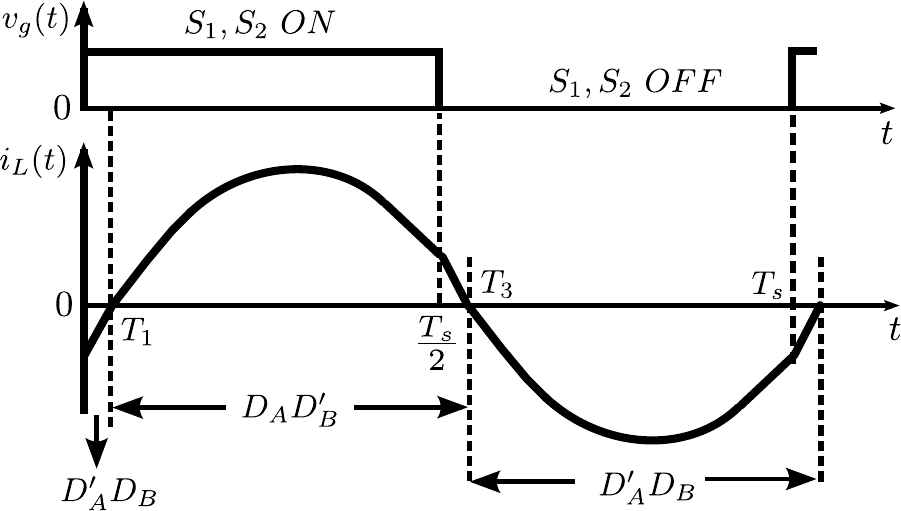}
\caption{Switching sequence of SRC.}
\label{Fig2}
\end{figure}
\section{Discrete State space model of SRC}
The SRC circuit with the labelled components are shown in Fig. \ref{Fig1a}. The state variables chosen are resonant capacitor ($C_r$) voltage $v_c(t)$, resonant inductor ($L_r$) current $i_L(t)$ and output capacitor ($C_o$) voltage $v_o(t)$. Turning ON of switch $S_1$ and $S_2$ at $t=0$ indicates the starting of first switch configuration in the operation of SRC out of the four switch configurations present in $T_s$ shown in Fig.~\ref{Fig2}. First switch configuration ends and second switch configuration starts when $i_L(t)$ reaches zero from negative polarity at time $T_1$. When switch $S_1$ and $S_2$ are turned OFF at time $T_s/2$ second switch configuration ends and third switch configuration starts~\cite{ref7}. Third switch configuration ends at time $T_3$ when $i_L(t)$ reaches zero from positive polarity and fourth switch configuration continues until $S_1$ and $S_2$ are again turned ON. The range of input ripple frequency $f_{in}$ is chosen from $100Hz$ to $10kHz$. The operating frequency of SRC, $f_{s}$, considered in the analysis is above its tank resonance frequency and are in the range of $100kHz$. Table~\ref{SRCparameter} gives the values of various SRC parameters considered for the time domain simulation for verifying the analysis.
\begin{table}[!t]
\renewcommand{\arraystretch}{1.3}
\caption{Nominal values of SRC parameters}
\label{SRCparameter}
\centering
\begin{tabular}{|l|r|}
\hline
\bfseries Parameters & \bfseries Values\\
\hline
DC input voltage ($V_{in}$) & $700V$\\
\hline
DC input ripple frequency & $100Hz - 10kHz$\\
\hline
Output power & $10kW$\\
\hline
DC output voltage ($V_o$) & $10kV$\\
\hline
DC output voltage ripple & $0.1\%$\\
\hline
SRC resonance frequency & $100kHz$\\
\hline
Range of $F$ & $1.01-1.5$\\
\hline
Range of $Q_e$ & $0.5-10$\\
\hline
\end{tabular}
\end{table}
\subsection{Exact discretization}
In this analysis the state space model of the SRC is derived using exact discretization modelling~\cite{ref4}. It is assumed that the states of the system are represented by linear time invariant differential equations and the state variable values are continuous from one switch configuration to the next. Hence the states at the end of one switch configuration form the initial condition for the next switch configuration. If $x\left[kT_s\right]$ is the initial condition of the state at $k^{th}$ sample and $u(t)$ is the input between $k^{th}$ sample and time $kT_s+t$, then states at $kT_s+t$ is,
\begin{equation} \label{sdm}
\begin{split}
x\left[kT_s+t\right]=&e^{A\left(\left[kT_s+t\right]-kT_s\right)}x\left[kT_s\right]+\\
&\int\limits_{kT_s}^{kT_s+t}e^{A\left(\left[kT_s+t\right]-\tau\right)}Bu(\tau)d\tau
\end{split}
\end{equation}
where, $A$, $B$ are the state matrices derived from dynamic equations valid between $kT_s$ and $kT_s+t$~\cite{ref4}. Initial conditions of the states $x\left[kT_s\right]=\left[I_{L}~V_{c}~V_{o}\right]^T$, where $I_L$, $V_c$ and $V_o$ are the values of the states at time $t=kT_s$. The input to the SRC at $t=kT_s$ is $u\left[kT_s\right]=\left[V_{in}\right]$. The term $e^{At}$ is computed by evaluating $L^{-1}\left(\left[sI-A\right]^{-1}\right)$.
\subsection{Non-linear state equations of SRC}
The state equations of SRC are formulated with $3$ state variables for the analysis of audiosusceptibility. This is carried out in two steps by using $2$ plus $1$ state variable approach. In \textit{Step 1}, $2$ state variables and in \textit{Step 2} remaining one state variable are considered.
\begin{figure}[!t]
\centering
\subfloat[]{\includegraphics[keepaspectratio,scale=0.63]{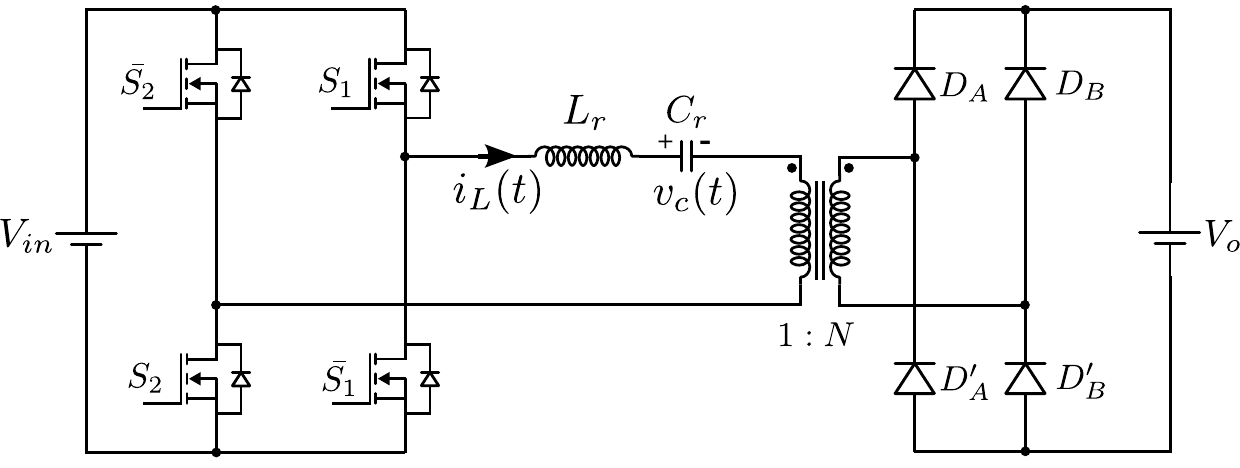}}\\
\subfloat[]{\includegraphics[keepaspectratio,scale=0.63]{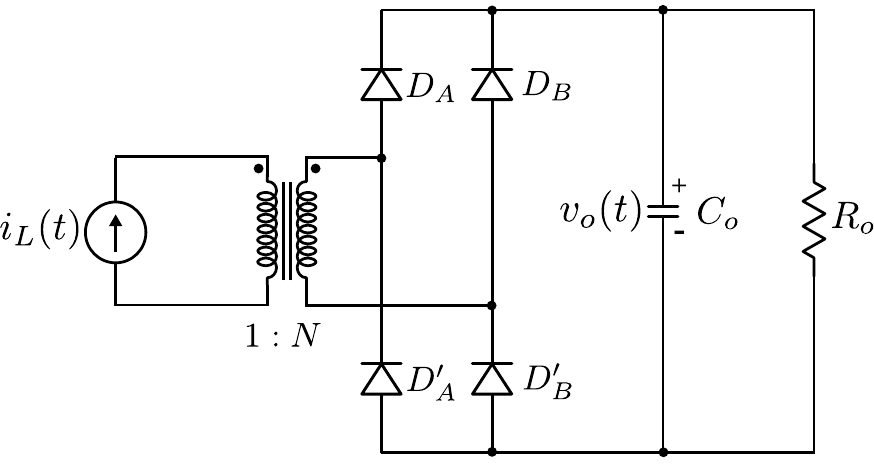}}
\caption{Equivalent circuit diagram with (a) constant dc output (b) input modelled as current source.} 
\label{Fig1b}
\end{figure}
\subsubsection{Step 1}
Since the range of $f_{in}$ of interest is an order less than $f_{s}$, the variation in $v_{in}(t)$ and $v_{o}(t)$ at frequency $f_{in}$ are neglected in a switching period $T_s$. This simplifies the analysis to $2$ state variables as shown in Fig.~\ref{Fig1b}(a). $x\left(t\right)=\left[i_L(t)~v_c(t)\right]^T$ with input $u\left(t\right)=\left[V_{in}~V_{o}\right]^T$ is evaluated and the state variable $\left[v_o(t)\right]$ is evaluated subsequently. The dynamic equations from the simplified equivalent diagram shown in Fig. \ref{Fig1b}(a) gives the state matrices related to the four switch configurations represented by suffix from $1$ to $4$ given by,
\begin{equation}
\setlength{\arraycolsep}{2pt}
{A}_1{=}A_2{=}A_3{=}A_4{=}
  \begin{bmatrix}
{0}&{-\cfrac{1}{L_r}}\\
{\cfrac{1}{C_r}}&{0}
  \end{bmatrix},
{B}_1{=}-B_3{=}
  \begin{bmatrix}
{\cfrac{1}{L_r}}&{\cfrac{1}{NL_r}}\\
{0}&{0}
  \end{bmatrix}\label{AB1}
\end{equation}
\begin{equation}
\setlength{\arraycolsep}{2pt}
{B}_2{=}-B_4{=}
  \begin{bmatrix}
{\cfrac{1}{L_r}}&{-\cfrac{1}{NL_r}}\\
{0}&{0}
  \end{bmatrix}\nonumber
\end{equation}
Substituting $\left(A_1,B_1\right)$ from (\ref{AB1}) and input $u\left(t\right)=\left[V_{in}~V_{o}\right]^T$ in (\ref{sdm}), the state at the end of first switch configuration $x\left[kT_s+T_1\right]$ at $t=T_1$ is derived which is the initial condition for the second switch configuration. The state at the end of second switch configuration $x\left[kT_s+(T_s/2)\right]$ is derived using the state matrices $\left(A_2,B_2\right)$ given in (\ref{AB1}) and the same input $u\left(t\right)$. This procedure is continued until the states at the end of switching period $x\left[kT_s+T_s\right]$ are derived.
\begin{equation}\label{x1}
\setlength{\arraycolsep}{2pt}
\begin{split}
&x\left[kT_s+T_s\right]=
  \begin{bmatrix}
\mathbf{a_{11}}&\mathbf{a_{12}}\\
\mathbf{a_{21}}&\mathbf{a_{22}}
  \end{bmatrix}x\left[kT_s\right]+
  \begin{bmatrix}
\mathbf{a_{13}}&\mathbf{b_{11}}\\
\mathbf{a_{23}}&\mathbf{b_{21}}
  \end{bmatrix}u\left[kT_s\right]\\
&\text{where}~~
x\left[kT_s\right]=
  \begin{bmatrix}
I_L~V_c
  \end{bmatrix}^T,\quad
u\left[kT_s\right]=
  \begin{bmatrix}
V_{o}&V_{in}
  \end{bmatrix}^T
\end{split}
\end{equation}
The elements $\mathbf{a_{13}}$ and $\mathbf{a_{23}}$ are defined as in (\ref{x1}) to aid the development of the state space model of the overall system as explained in section~\ref{Aelement}. The elements in the matrices are given in Appendix \ref{Ad}.
\subsubsection{Step 2}
$v_o(t)$ is determined by the current $i_L(t)$. Hence, $i_L(t)$ is considered as an input current source for the estimation of the state variable $v_o(t)$ for the period $T_s$ where $i_L(t)$ is updated in \textit{Step 1} in each $T_s$. This simplifies the circuit to one state variable $x\left(t\right)=\left[v_o(t)\right]$ and input $u\left(t\right)=\left[i_L(t)\right]$. The state matrices related to four switch configurations represented by suffix from $1$ to $4$ are obtained using the dynamic equations from the simplified equivalent diagram shown in Fig. \ref{Fig1b}(b).
\begin{equation}\label{AB2}
\begin{split}
&{A^\prime_1}{=}A^\prime_2{=}A^\prime_3{=}A^\prime_4{=}
  \begin{bmatrix}
{-\cfrac{1}{R_oC_o}}
  \end{bmatrix},\\
&{B^\prime_1}{=}-B^\prime_2{=}-B^\prime_3{=}B^\prime_4{=}
  \begin{bmatrix}
{-\cfrac{1}{NC_o}}
  \end{bmatrix}
\end{split}
\end{equation}
The input $u\left(t\right)$ for each switch configuration can be found from the solution of the state $i_L\left(t\right)$ for each switch configuration derived in \textit{Step 1}. Substituting $(A^\prime,B^\prime)$ matrices given in (\ref{AB2}) and $u\left(t\right)$ obtained from \textit{Step 1} in (\ref{sdm}), the states for each switch configuration are derived. Using the state at the end of each switch configuration as an initial condition for the next switch configuration, the state at the end of $T_s$ is given by,
\begin{equation}\label{x2}
\begin{split}
&x\left[kT_s+T_s\right]=
  \begin{bmatrix}
\mathbf{a_{33}}
  \end{bmatrix}x\left[kT_s\right]+
  \begin{bmatrix}
\mathbf{a_{31}}&\mathbf{a_{32}}&\mathbf{b_{31}}
  \end{bmatrix}u\left[kT_s\right]\\
&\text{where}~x\left[kT_s\right]=
  \begin{bmatrix}
V_o
  \end{bmatrix},\quad
u\left[kT_s\right]=
  \begin{bmatrix}
I_{L}&V_c&V_{in}
  \end{bmatrix}^T
\end{split}
\end{equation}
The elements in the matrices of (\ref{x2}) are given in Appendix \ref{Bd}. Even though the input $u(t)=\left[i_L(t)\right]$, the waveform for $i_L(t)$ also depends on the initial conditions of $v_c(t)$ and $v_{in}(t)$. Hence, this results in $u\left[kT_s\right]=\left[I_{L}~V_c~V_{in}\right]^T$ as expressed in (\ref{x2}).
\setlength{\arraycolsep}{5pt}
\subsection{Discrete state space model of SRC}\label{Aelement}
The discrete domain state space model is obtained by combining the $2$ state variables model in \textit{step1} and $1$ state variable model in \textit{step2} given in (\ref{x1}) and (\ref{x2}) respectively. This yields the complete discrete domain state space model of the SRC.
\begin{equation}\label{x}
x\left[kT_s+T_s\right]=A_{d}x\left[kT_s\right]+B_{d}u\left[kT_s\right]
\end{equation}
\begin{equation}\label{xu}
\begin{split}
&A_{d}{=}
  \begin{bmatrix}
\mathbf{a_{11}}&\mathbf{a_{12}}&\mathbf{a_{13}}\\
\mathbf{a_{21}}&\mathbf{a_{22}}&\mathbf{a_{23}}\\
\mathbf{a_{31}}&\mathbf{a_{32}}&\mathbf{a_{33}}
  \end{bmatrix},\quad
B_{d}{=}
  \begin{bmatrix}
\mathbf{b_{11}}\\
\mathbf{b_{21}}\\
\mathbf{b_{31}}
  \end{bmatrix},\\
&x\left[kT_s\right]=
  \begin{bmatrix}
I_L&V_c&V_o
  \end{bmatrix}^T,\quad
u\left[kT_s\right]=
  \begin{bmatrix}
V_{in}
  \end{bmatrix}
\end{split}
\end{equation}
Using (\ref{x}) the SRC is described by the states, input and the time durations as shown in Fig.~\ref{Fig2}. Hence (\ref{x}) can be described by,
\begin{equation}\label{xm}
x\left[kT_s+T_s\right]=f\left(x,u,T_s,T_1,T_3\right)
\end{equation}
This is a state space model of a discrete-time non-linear system. It can be linearized after identifying the equilibrium operating trajectory for the system.
%
%
\section{Small signal model of SRC}\label{sec3}
The steady state waveforms of the SRC are periodic with period $T_s$. The steady state values of the $x\left[kT_s\right]$ are derived as follows.
\subsection{Cyclic steady state model of SRC}
The cyclic steady state variables of the SRC is denoted by the superscript $s$ if (\ref{T1a}) and (\ref{T1}) are satisfied~\cite{ref4}.
\setlength{\arraycolsep}{0.10em}
\begin{equation}\label{T1a}
x^s\left[kT_s+T_s\right]=f\left(x^s,u^s,T^s_s,T^s_1,T^s_3\right){=}x^s\left[kT_s\right]
\end{equation}
\begin{equation}\label{T1}
i^s_{L}(kT_s+T_1)=0,\quad i^s_L(kT_s+T_3)=0,
\end{equation}
\setlength{\arraycolsep}{5pt}
Equation (\ref{T1a}) represents the cyclic steady state condition of the states and (\ref{T1}) represents the constraint for times $T_1$ and $T_3$. Substituting the state equation for $i_L(t)$ from (\ref{x}) in (\ref{T1}), the cyclic steady state time $T^s_1$ and $T^s_3$ is constrained by (\ref{T1s}) and (\ref{T3s}).
\begin{align}
&f_{T1}=I_L^s\cos\omega_rT_1^s-\dfrac{V_c^s-V_{in}^s-\tfrac{V_o^s}{N}}{Z_c}\sin\omega_rT_1^s=0\label{T1s}\\
\begin{split}
&f_{T3}=I_L^s\cos\omega_rT_3^s-\dfrac{V_c^s-V_{in}^s-\tfrac{V_o^s}{N}}{Z_c}\sin\omega_rT_3^s-\\
&\dfrac{2V_o^s}{NZ_c}\sin\omega_r(T_3^s-T_1^s)-\dfrac{2V_{in}^s}{Z_c}\sin\omega_r(T_3^s-\tfrac{T_s}{2})=0
\end{split}\label{T3s}
\end{align}
\subsection{Small signal model of SRC}
Since the small signal analysis is restricted to audiosusceptibility the switching period $T_s$ of SRC is held constant and a small variation for the input is applied about the cyclic steady state. This resulting variations are denoted by the symbol $\tilde{}$ in the states, input, and subinterval times and are given by (\ref{xut_t}).
\setlength{\arraycolsep}{0.10em}
\begin{equation}\label{xut_t}
\begin{split}
&\tilde{i}_L\left[kT_s\right]{=}I_L\left[kT_s\right]{-}I^s_L\left[kT_s\right],\\
&\tilde{v}_c\left[kT_s\right]{=}V_c\left[kT_s\right]{-}V^s_c\left[kT_s\right],\\
&\tilde{v}_o\left[kT_s\right]{=}V_o\left[kT_s\right]{-}V^s_o\left[kT_s\right],\\
&\tilde{v}_{in}\left[kT_s\right]{=}V_{in}\left[kT_s\right]{-}V^s_{in}\left[kT_s\right],\\
&\tilde{t}_{1,k}{=}T_{1,k}{-}T^s_{1,k},\quad\tilde{t}_{3,k}{=}T_{3,k}{-}T^s_{3,k}
\end{split}
\end{equation}
where, $T_{1,k}$ and $T_{3,k}$ are the values of $T_1$ and $T_3$ during the $k^{th}$ sample.

By substituting (\ref{xut_t}) in (\ref{x}) the variation in states at $kT_s+T_s$ is,
\begin{eqnarray}
\tilde{x}\left[kT_s+T_s\right]&=&x\left[kT_s+T_s\right]-x^s\left[kT_s+T_s\right]
\end{eqnarray}
Applying the Taylor series expansion at the steady-state operating point in (\ref{xm}) and truncating the higher order non-linear terms gives the linearized small signal model as,
\begin{eqnarray}\label{pd}
\setlength{\arraycolsep}{0.0em}
\tilde{x}\left[kT_s{+}T_s\right]{=}{{\left.\dfrac{\partial f}{\partial x}\right\rvert_{x^s}}{\tilde{x}}}{+}{{\left.\dfrac{\partial f}{\partial u}\right\rvert_{u^s}}{\tilde{u}}}{+}{{\left.\dfrac{\partial f}{\partial T_1}\right\rvert_{T_1^s}}{\tilde{t}_1}}{+}{{\left.\dfrac{\partial f}{\partial T_3}\right\rvert_{T_3^s}}{\tilde{t}_3}}~~~~~
\end{eqnarray}
\setlength{\arraycolsep}{5pt}
Since the elements of matrices $A_d$ and $B_d$ are functions of time $T_1$ and $T_3$, substituting (\ref{x}) in (\ref{pd}) gives,
\begin{eqnarray}\label{ss1}
\tilde{x}\left[kT_s+T_s\right]=A_d\tilde{x}\left[kT_s\right]+B_d\tilde{u}\left[kT_s\right]+T_d\tilde{t}_k
\end{eqnarray}
where,
\begin{equation}\label{Tds}
\setlength{\arraycolsep}{0.50em}
\begin{split}
&T_d{=}
{\begin{bmatrix}
\setlength{\arraycolsep}{0.0em}
  {\begin{bmatrix}
  \setlength{\arraycolsep}{0.10em}
  \dfrac{\partial A_d}{\partial T_1}&\dfrac{\partial B_d}{\partial T_1}
  \end{bmatrix}
  \begin{bmatrix}
  \setlength{\arraycolsep}{0.0em}
  x\left[kT_s\right]\\u\left[kT_s\right]
  \end{bmatrix}}
&&
  \setlength{\arraycolsep}{0.0em}
  {\begin{bmatrix}
\setlength{\arraycolsep}{0.0em}
  \dfrac{\partial A_d}{\partial T_3}&\dfrac{\partial B_d}{\partial T_3}
  \end{bmatrix}
  \begin{bmatrix}
  \setlength{\arraycolsep}{0.10em}
  x\left[kT_s\right]\\u\left[kT_s\right]
  \end{bmatrix}}
\end{bmatrix}},\\
&{\tilde{x}\left[kT_s\right]{=}
  \begin{bmatrix}
  \tilde{i}_L\\
  \tilde{v}_c\\
  \tilde{v}_o
  \end{bmatrix},\quad
  \tilde{u}\left[kT_s\right]{=}
    \begin{bmatrix}
    \tilde{v}_{in}
    \end{bmatrix},\quad
\tilde{t}_k{=}
  \begin{bmatrix}
  \tilde{t}_1\\
  \tilde{t}_3
  \end{bmatrix}}
\end{split}
\end{equation}
\begin{figure*}[!b]
\setcounter{MYtempeqncnt}{\value{equation}}
\setcounter{equation}{22}
\hrulefill
\begin{align}
&\dfrac{v_o\left(z\right)}{v_{in}\left(z\right)}=\dfrac{\dfrac{16}{NZ_cC_o\omega_r}\biggl\{\left(z-1\right)-\dfrac{4V_o}{NZ_cf^{\prime}_{T1}}\biggr\}}{\left(z-1\right)^3-\dfrac{4V_o}{NZ_cf^{\prime}_{T1}}\left(z-1\right)^2+\dfrac{16}{N^2Z_cC_o\omega_r}\left(z-1\right)-\dfrac{64V_o}{N^3Z^2_cC_o\omega_rf^{\prime}_{T1}}}\label{TF}
\end{align}
\setcounter{equation}{\value{MYtempeqncnt}}
\end{figure*}
The order of $T_d$ is $3{\times}2$ where the columns are obtained by the product of two matrices of the order $3{\times}4$ and $4{\times}1$. Using (\ref{xu}) in the expression for $T_d$ in (\ref{Tds}), the elements $\mathbf{td_{pq}}$ of $T_d$, where suffix $(p,q)$ represent the row and column indices, are tabulated in Appendix \ref{td}.
\setlength{\arraycolsep}{0.0em}
\setlength{\arraycolsep}{5pt}
\vspace{0pt}
\subsubsection{Estimation of $\tilde{t}_k$}
Expanding the dynamics constraint in (\ref{T1s}) and (\ref{T3s}) by using Taylor series expansion and truncating the higher order non-linear terms gives,
\begin{small}
\begin{align}
\label{pd1}
&\tilde{f}_{T1}{=}{{\left.\dfrac{\partial f_{T1}}{\partial x}\right\vert_{x^s}}{\tilde{x}}}{+}{{\left.\dfrac{\partial f_{T1}}{\partial u}\right\vert_{u^s}}{\tilde{u}}}{+}{{\left.\dfrac{\partial f_{T1}}{\partial T_1}\right\vert_{T_1^s}}{\tilde{t}_1}}{+}{{\left.\dfrac{\partial f_{T1}}{\partial T_3}\right\vert_{T_3^s}}{\tilde{t}_3}}{=}0\\
\label{pd2}
&\tilde{f}_{T3}{=}{{\left.\dfrac{\partial f_{T3}}{\partial x}\right\vert_{x^s}}{\tilde{x}}}{+}{{\left.\dfrac{\partial f_{T3}}{\partial u}\right\vert_{u^s}}{\tilde{u}}}{+}{{\left.\dfrac{\partial f_{T3}}{\partial T_1}\right\vert_{T_1^s}}{\tilde{t}_1}}{+}{{\left.\dfrac{\partial f_{T3}}{\partial T_3}\right\vert_{T_3^s}}{\tilde{t}_3}}{=}0
\end{align}
\end{small}
By solving (\ref{T1s}) and (\ref{T3s}), $\tilde{t}_1$ and $\tilde{t}_3$ can be represented in terms of $\tilde{x}$ and $\tilde{u}$ as,
\begin{eqnarray}\label{pdt1t3}
\tilde{t}_{k}{=}T_{kx}\tilde{x}\left[kT_s\right]+T_{ku}\tilde{u}\left[kT_s\right]
\end{eqnarray}
The elements of $T_{kx}$ and $T_{ku}$ are denoted by $\mathbf{tx_{pq}}$ and $\mathbf{tu_{pq}}$  respectively where the suffix represents the row and column numbers. These elements are given in Appendix \ref{tk}.
Substituting (\ref{pdt1t3}) in (\ref{ss1}) gives the overall small signal model of SRC in terms of states and input and is given by,
\begin{equation}\label{ssf}
\begin{split}
&\tilde{x}\left[kT_s+T_s\right]=A_{sd}\tilde{x}\left[kT_s\right]+B_{sd}\tilde{u}\left[kT_s\right]\\
&\text{where}\quad A_{sd}=\left[A_d+T_dT_{kx}\right]\quad B_{sd}=\left[B_d+T_dT_{ku}\right]
\end{split}
\end{equation}
This represents a system with states representing the incremental components of resonant capacitor voltage, resonant inductor current and output voltage at the discrete time sampling time instants. The input for this model is the incremental input voltage, $\tilde{v}_{in}\left[kT_s\right]$.
\subsection{Simplified small signal model of SRC}
The dynamic model in (\ref{ssf}) can be used to obtain the relationships between the input voltage $\tilde{v}_{in}\left[kT_s\right]$ and the state variables $\tilde{x}\left[kT_s\right]$ with the coefficients of the dynamic system matrices given in the Appendix. However, it is possible to simplify the entries of the matrices by making some practical assumptions.

The major assumption made for simplifying the small signal model is $T_3=(T_s/2)+T_1$. Additional practical assumptions made are $e^{\tfrac{t}{R_oC_o}}\approx 1,~\forall ~0\le t \le T_s$ and $\tfrac{1}{(R_oC_o)^2}\ll\omega_r^2$. This implies that the output voltage decay is neglected over a switching period, and the reciprocal of the output time constant is much lower than the tank resonance frequency. These simplify the $A,B$ matrices of the small signal model as,
\begin{equation}\label{AsBs}
\setlength{\arraycolsep}{2pt}
\overline{A}_{sd}{=}
  \begin{bmatrix}
1{+}\dfrac{4\omega_rV_o}{NZ_cf^{\prime}_{T1}} & -\dfrac{\sin\omega_rT_s}{Z_c} & 0 \\
Z_c\sin\omega_rT_s & 1 & \dfrac{4}{N} \\
0 & -\dfrac{4}{NZ_cC_o\omega_r} & 1 
  \end{bmatrix}{,}~
\overline{B}_{sd}{=}
  \begin{bmatrix}
0 \\
-4 \\
0 
  \end{bmatrix}
\setlength{\arraycolsep}{5pt}
\end{equation}
\addtocounter{equation}{1}
\begin{figure}[!t]
\centering
\subfloat[]{\includegraphics[keepaspectratio,scale=0.33]{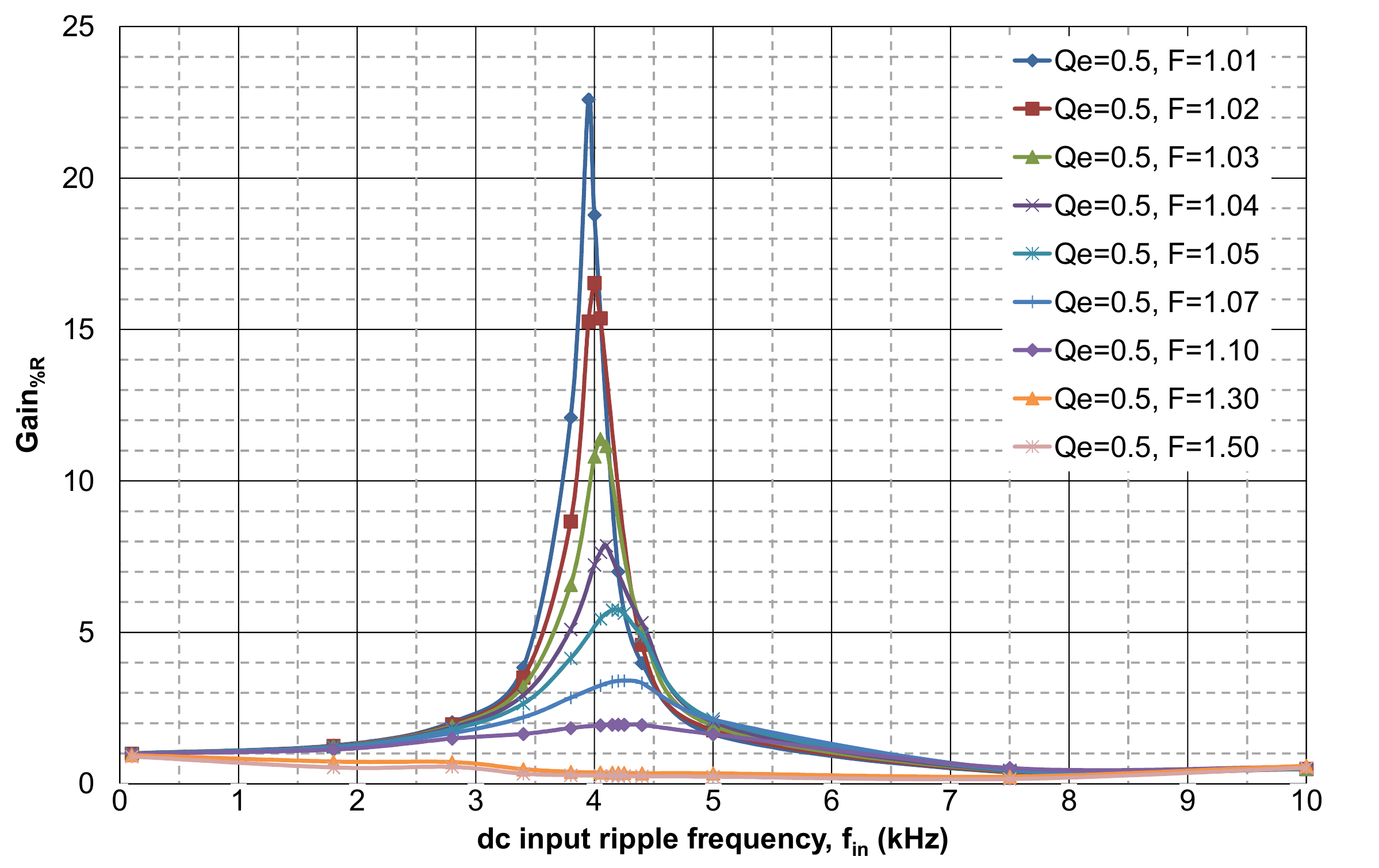}}\\
\subfloat[]{\includegraphics[keepaspectratio,scale=0.33]{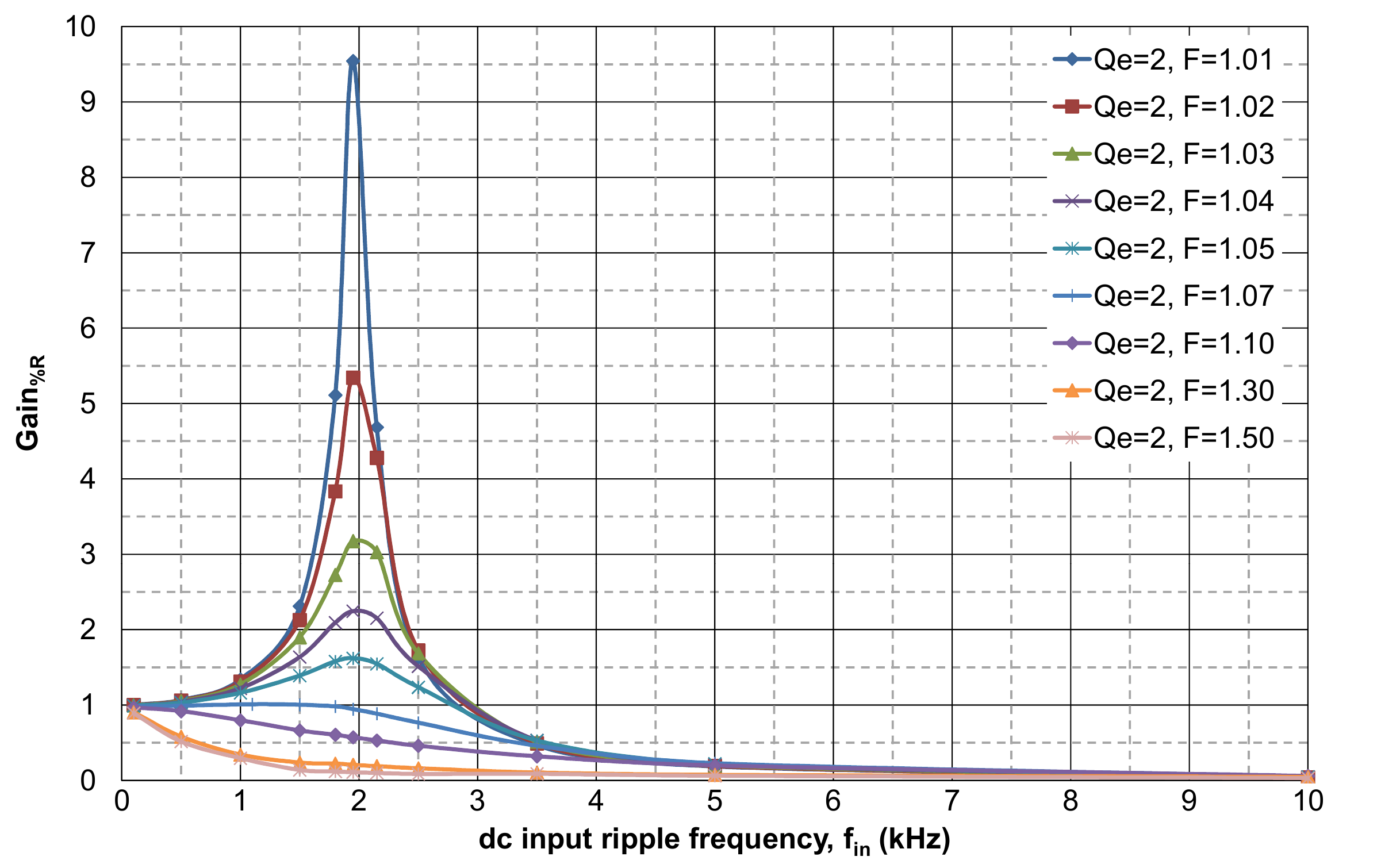}}
\caption{Gain versus $f_{in}$ from simulation for $F\in\left[1.01,1.5\right]$ and (a) $Q_e=0.5$ (b) $Q_e=2.0$.} 
\label{Fig3ab}
\end{figure}
\begin{figure}[!t]
\centering
\subfloat[]{\includegraphics[keepaspectratio,scale=0.33]{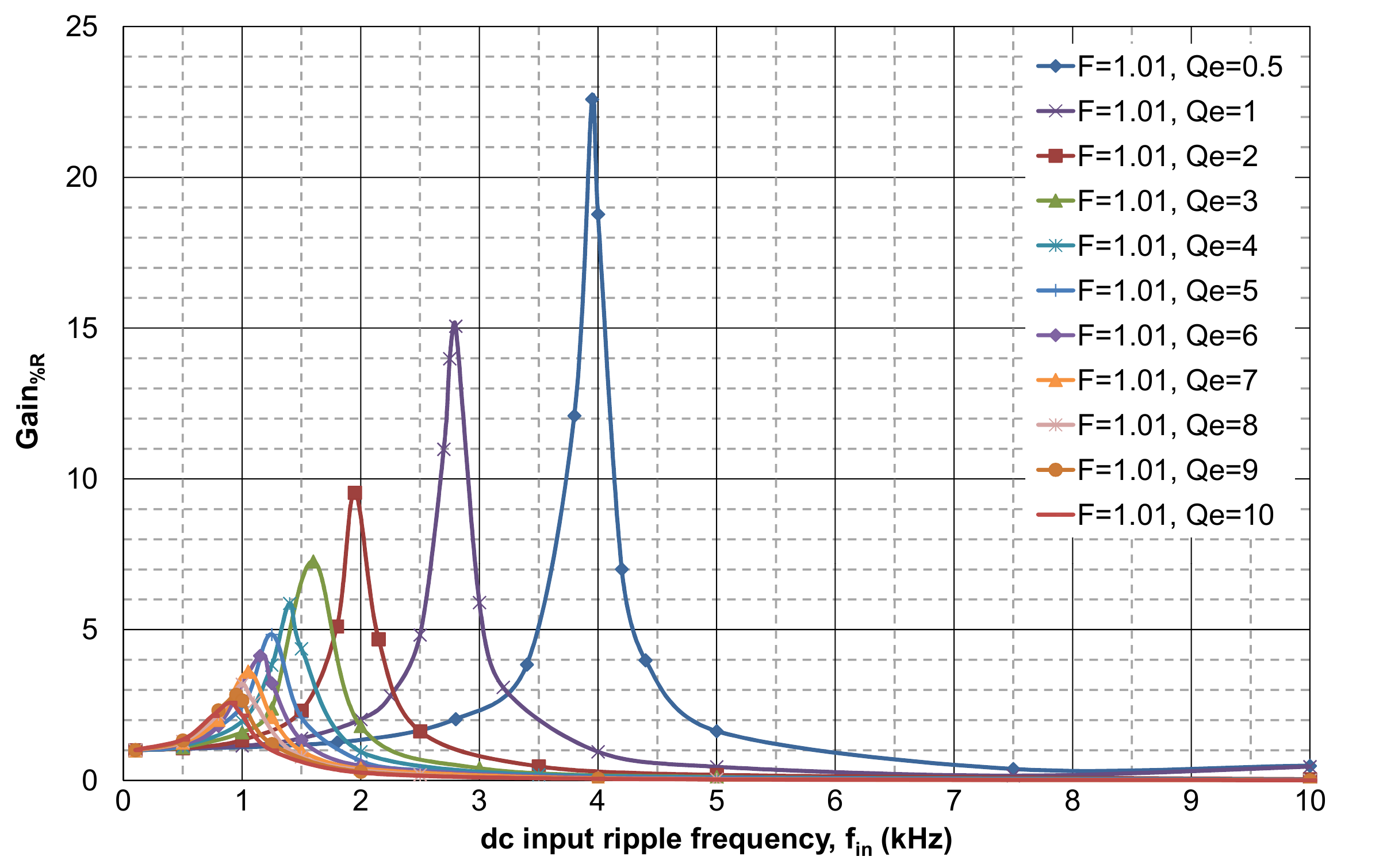}}\\
\subfloat[]{\includegraphics[keepaspectratio,scale=0.33]{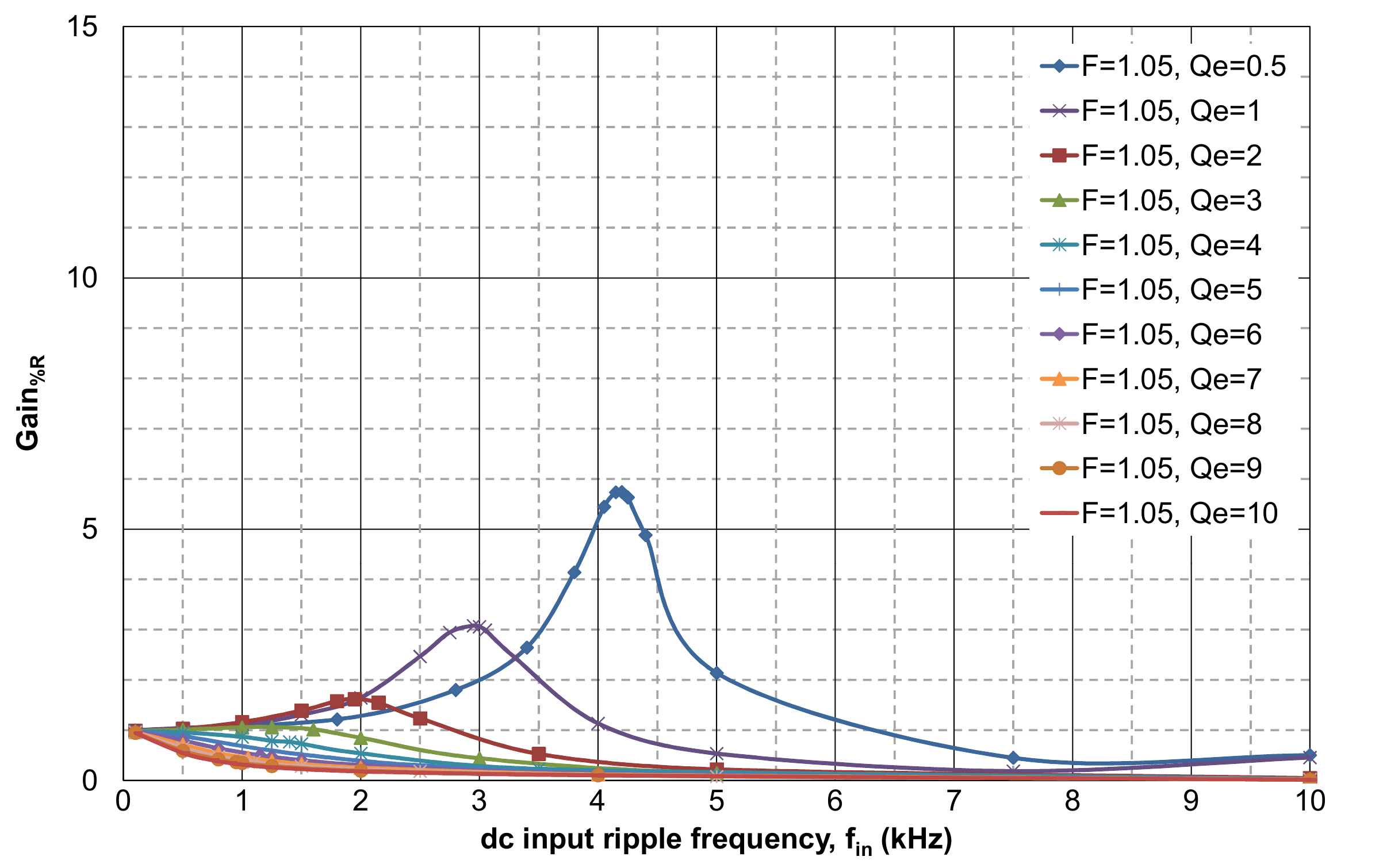}}
\caption{Gain versus $f_{in}$ from simulation for $Q_e\in\left[0.5,10\right]$ and (a) $F=1.01$ (b) $F=1.05$.} 
\label{Fig3bc}
\end{figure}
Small signal input to output transfer functions in the z-domain are found by substituting (\ref{AsBs}) in (\ref{ssf}). The obtained audiosusceptibility transfer function in z-domain is given in (\ref{TF}) where $f^{\prime}_{T1}{=}{-}df_{T1}/dt$.
The complex roots of denominator shows a resonance in the frequency response and the possibility of high gain of the SRC for input ripple. The s-domain AS resonance frequency for input ripple is obtained from the complex roots of denominator polynomial and is given by,
\begin{eqnarray}\label{winr}
\omega_{in,r}=\dfrac{1}{T_s}\tan^{-1}\left(\sqrt{\dfrac{16}{N^2C_o\omega_rZ_c}}\right)
\end{eqnarray}
The small signal analytical model for audiosusceptibility of a SRC given in (\ref{TF}) is compared with the time domain simulations and experimental results to verify the analytical model.
\begin{figure}[!t]
\centering
\subfloat[]{\includegraphics[keepaspectratio,scale=0.33]{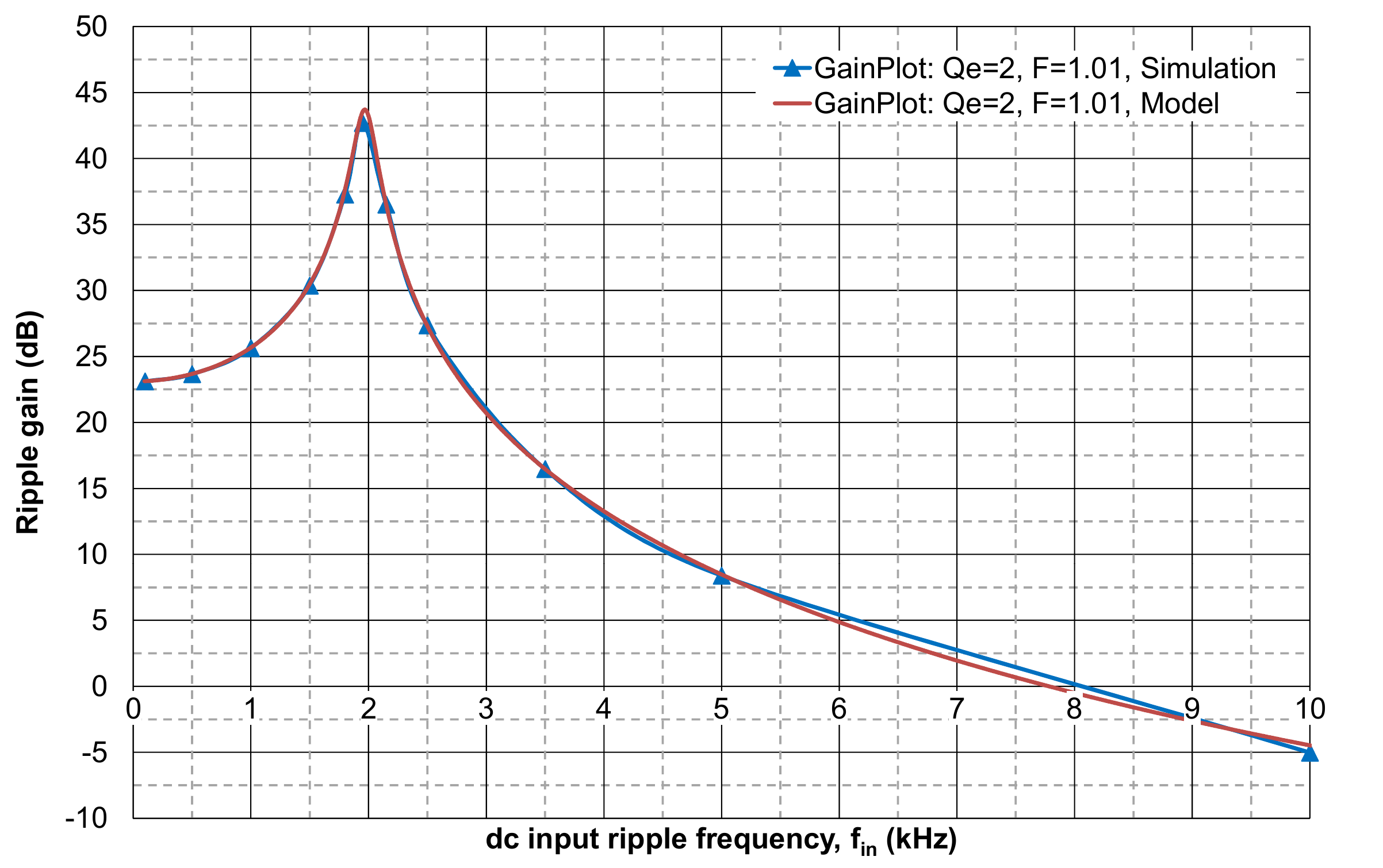}}\\
\subfloat[]{\includegraphics[keepaspectratio,scale=0.33]{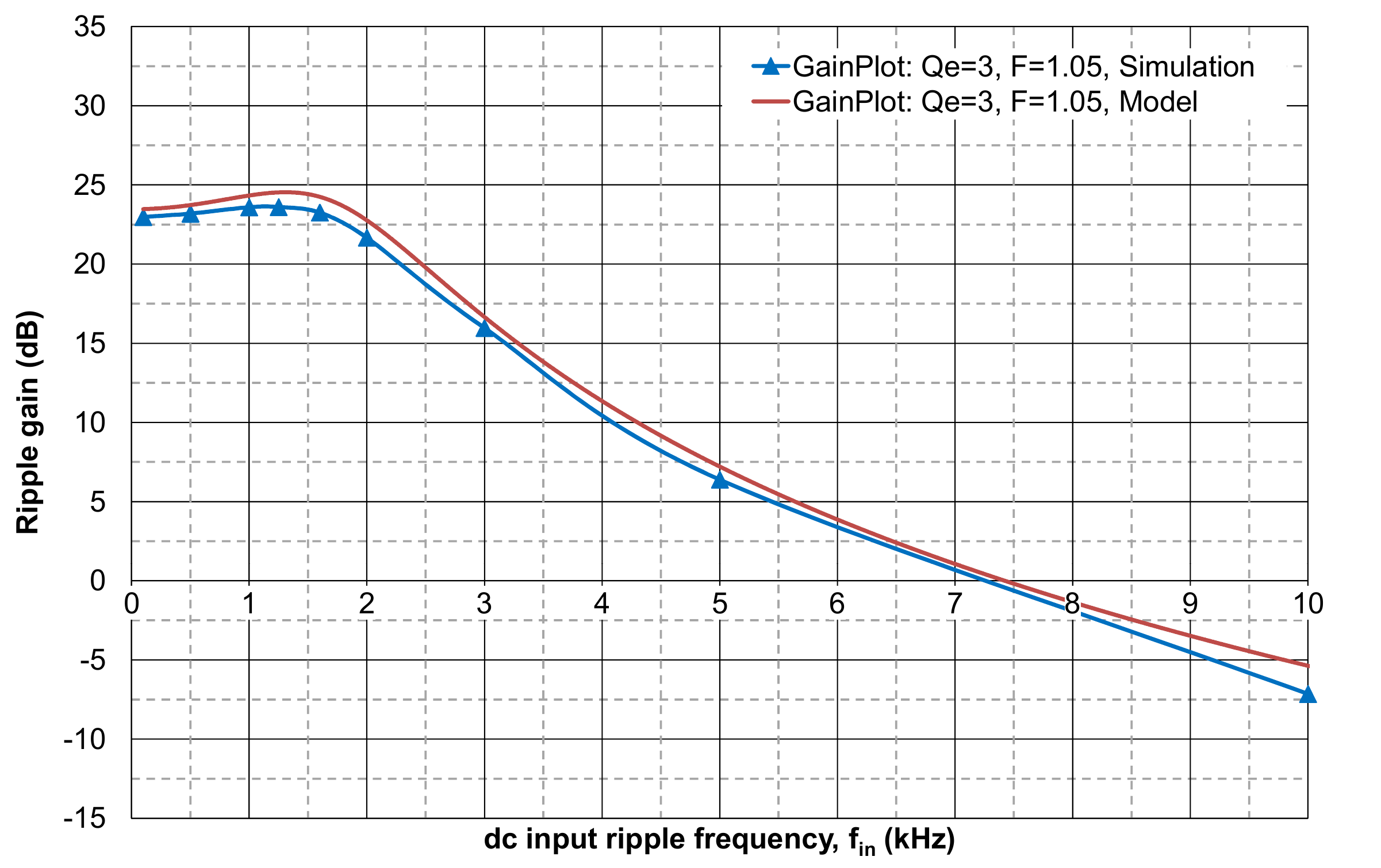}}
\caption{Comparison of small signal model with simulation for frequency response when (a) $Q_e=2$ and $F=1.01$ (b) $Q_e=3$ and $F=1.05$.} 
\label{Fig3cd}
\end{figure}
\begin{figure}[!t]
\centering
\subfloat[]{\includegraphics[keepaspectratio,scale=0.33]{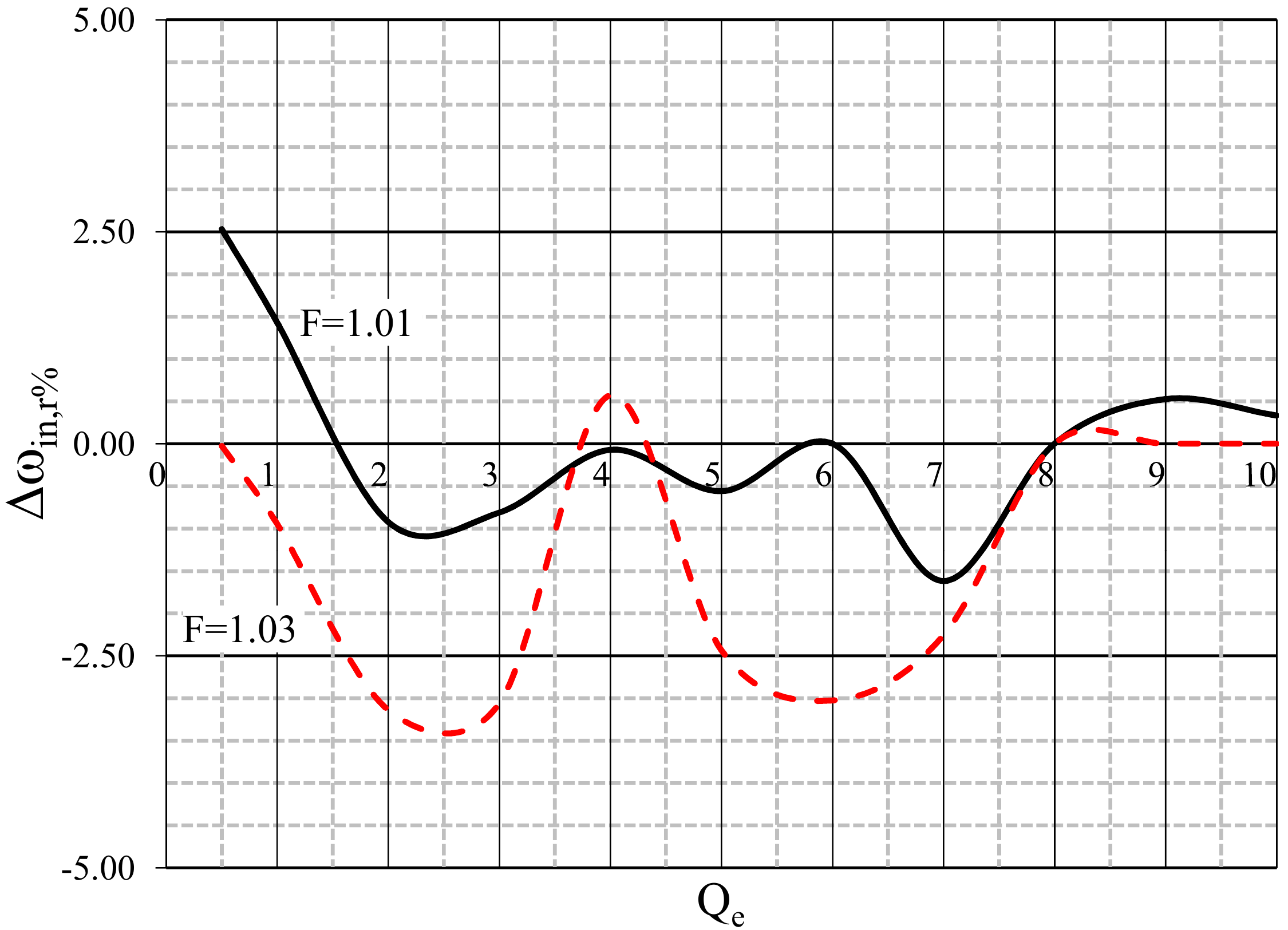}}\\
\subfloat[]{\includegraphics[keepaspectratio,scale=0.33]{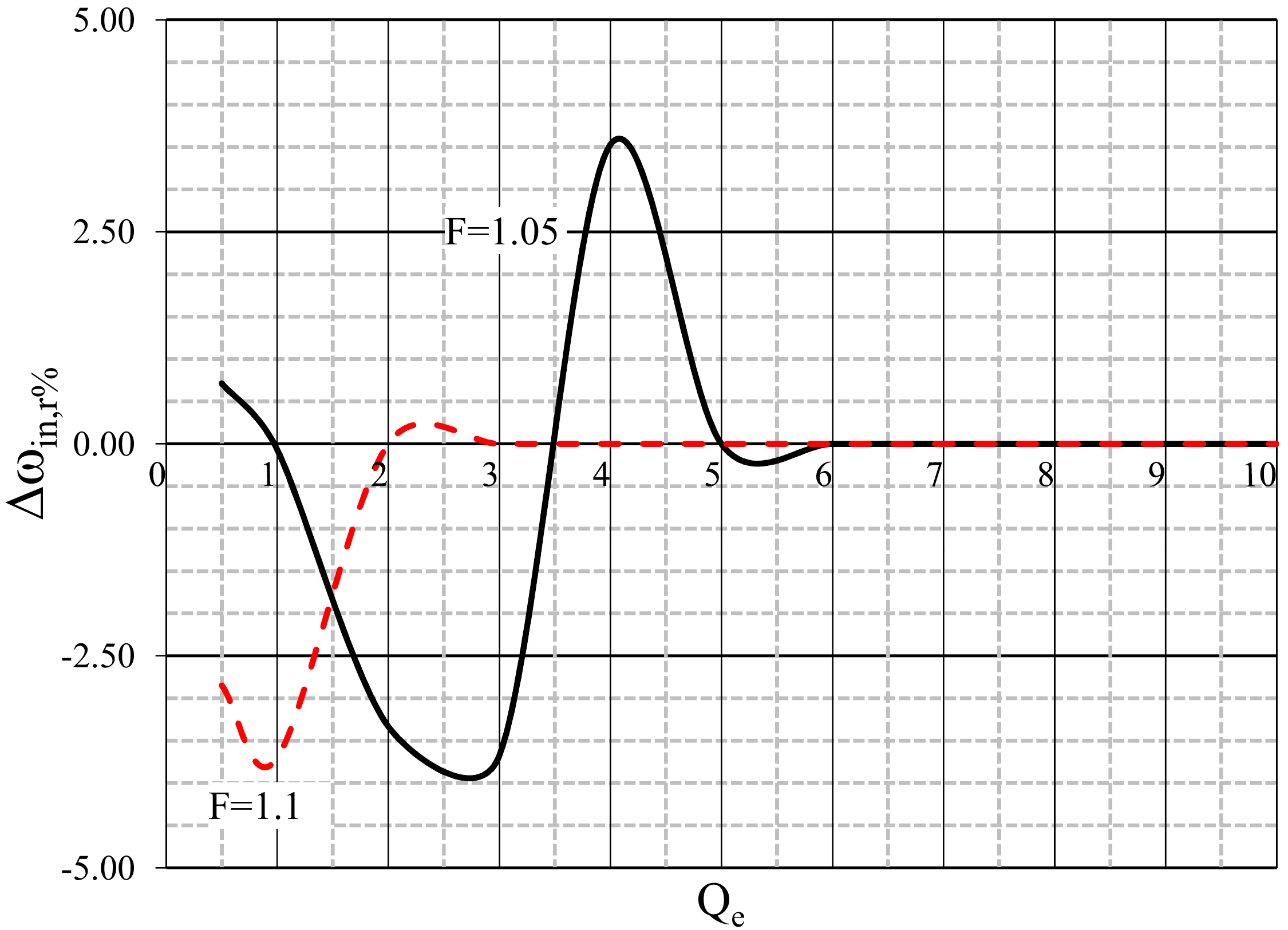}}
\caption{$\Delta\omega_{in,r\%}$ for $Q_e\in\left[0.5,10\right]$ and (a) for $F=1.01$ and $F=1.03$ (b) for $F=1.05$ and $F=1.1$.} 
\label{Fig3de}
\end{figure}
\section{Simulation results}
Conventionally, SRC design is based on two quantities $(i)~F$ defined as the ratio of switching frequency $f_s$ to tank resonance frequency $f_r$ of SRC $(ii)~Q_e$ defined as the effective quality factor of SRC~\cite{text1}. Hence to cover a wide range of SRC design, analysis is carried out based on $F$ and $Q_e$ where $F\in\left[1.01,1.5\right]$ and $Q_e\in\left[0.5,10\right]$. Time domain simulations are carried out for the validation of model in (\ref{TF}) where normalized gain is defined as, 
\begin{equation}\label{gainASp}
Gain_{\%R}=\dfrac{\vert\tilde{v}_{0}/\tilde{v}_{in}\vert}{\vert V_{0}/V_{in}\vert}
\end{equation}

In Figs.~\ref{Fig3ab}(a) and (b) the input ripple frequency is varied from $100Hz$ to $10kHz$ keeping the quality factor $Q_e{=}0.5$ and $Q_e{=}2.0$ respectively. At each frequency the gain, $Gain_{\%R}$, is computed from (\ref{gainASp}) for different values of $F\in\left[1.01,1.5\right]$. In both Figs.~\ref{Fig3ab}(a) and (b), the peak $Gain_{\%R}$ reduces as $F$ increases. Also from Fig.~\ref{Fig3ab}(a) the maximum $Gain_{\%R}$ is $23$ which occurs at $\omega_{in,r(s)}$ of $3950Hz$ for $F=1.01$. The decrease in $Gain_{\%R}$ for higher $F$ is found to be consistent for other $Q_e\in\left[0.5,10\right]$. Figs.~\ref{Fig3bc}(a) and (b) shows $Gain_{\%R}$ for $f_{in}$ varied from $100Hz$ to $10kHz$ keeping $F=1.01$ and $F=1.05$ respectively. The $Gain_{\%R}$ reduces as $Q_e$ increases in both Figs.~\ref{Fig3bc}(a) and (b). The decrease in $Gain_{\%R}$ for higher $Q_e$ is also found to be consistent for other $F\in\left[1.01,1.5\right]$.

The gain plot from simulation and analytical model in (\ref{TF}) are compared in Figs.~\ref{Fig3cd}(a) and (b) for $Q_e$ and $F$ of ($2$,$1.01$) and ($3$,$1.05$) respectively. In these gain plots the gain is defined as,
\begin{equation}
Ripple~gain=20\log_{10}(\vert\tilde{v}_o/\tilde{v}_{in}\vert)~~dB
\end{equation}
The Figs.~\ref{Fig3cd}(a) and (b) show that the simulation and the analytical models are closely matching.

$\omega_{in,r}$ and $\omega_{in,r(s)}$ are the frequencies at which AS resonance occurs in the small signal analytical model and in the time domain simulation respectively. Let the error in the frequency be defined as,
\begin{equation}\label{errw}
\Delta\omega_{in,r\%}=\dfrac{\omega_{in,r(s)}-\omega_{in,r}}{\omega_{in,r(s)}}100
\end{equation}
The $\Delta\omega_{in,r\%}$ computed for $F{=}1.01$, $F{=}1.03$ and $Q_e\in\left[0.5,10\right]$ is shown in Fig.~\ref{Fig3de}(a). Similar results are shown for $F{=}1.03$, $F{=}1.1$ and $Q_e\in\left[0.5,10\right]$ in Fig.~\ref{Fig3de}(b). Maximum error $\Delta\omega_{in,r\%}$ is found to be within $\pm 4\%$ which indicate good match of the analytical audiosusceptibility resonance frequency prediction with simulation results.
\subsection{SRC design for superior audiosusceptibility}
\begin{figure}[!t]
\centering
\includegraphics[keepaspectratio,scale=0.64]{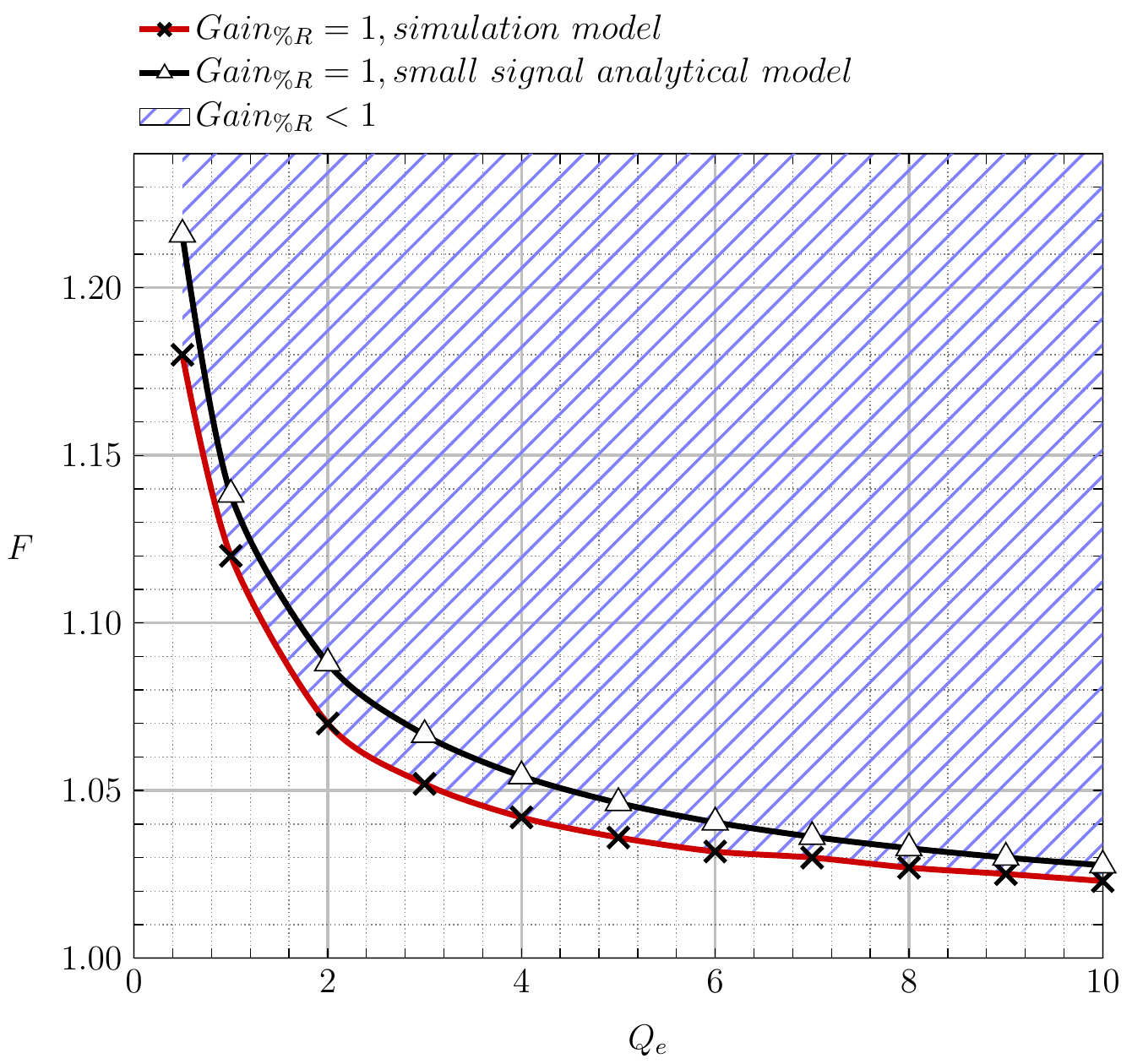}
\caption{Curve in the $F$ versus $Q_e$ plain for which audiosusceptibility gain is unity. For the region above the curve $Gain_{\%R}$ is less than unity. The curves from the small signal model and simulation are compared.}
\label{Fig3e}
\end{figure}

Each curve in Figs.~\ref{Fig3ab}(a) and (b) and Figs.~\ref{Fig3bc}(a) and (b) represents an SRC design with appropriated $F$ and $Q_e$ parameters. From the plots it is seen that there exists a set of values for $F$ and $Q_e$ where audiosusceptibility gain of the SRC is always less than unity for any $f_{in}$. Such values are identified and plotted as a region above the curves shown in Fig.~\ref{Fig3e}. Such a region obtained based on the analytical model using the gain expression in (\ref{TF}) and that using simulations are also compared in Fig. \ref{Fig3e}. Design of the SRC using $F$ and $Q_e$ values from this region is advantageous as it will have reduced audiosusceptibility with inherent ability to reject disturbance in the input voltage, $V_{in}$.
%
%
\section{Experimental results}\label{7.6}
\subsection{Test setup}
\begin{figure}[!t]
\centering
\subfloat[]{\includegraphics[keepaspectratio,scale=0.55]{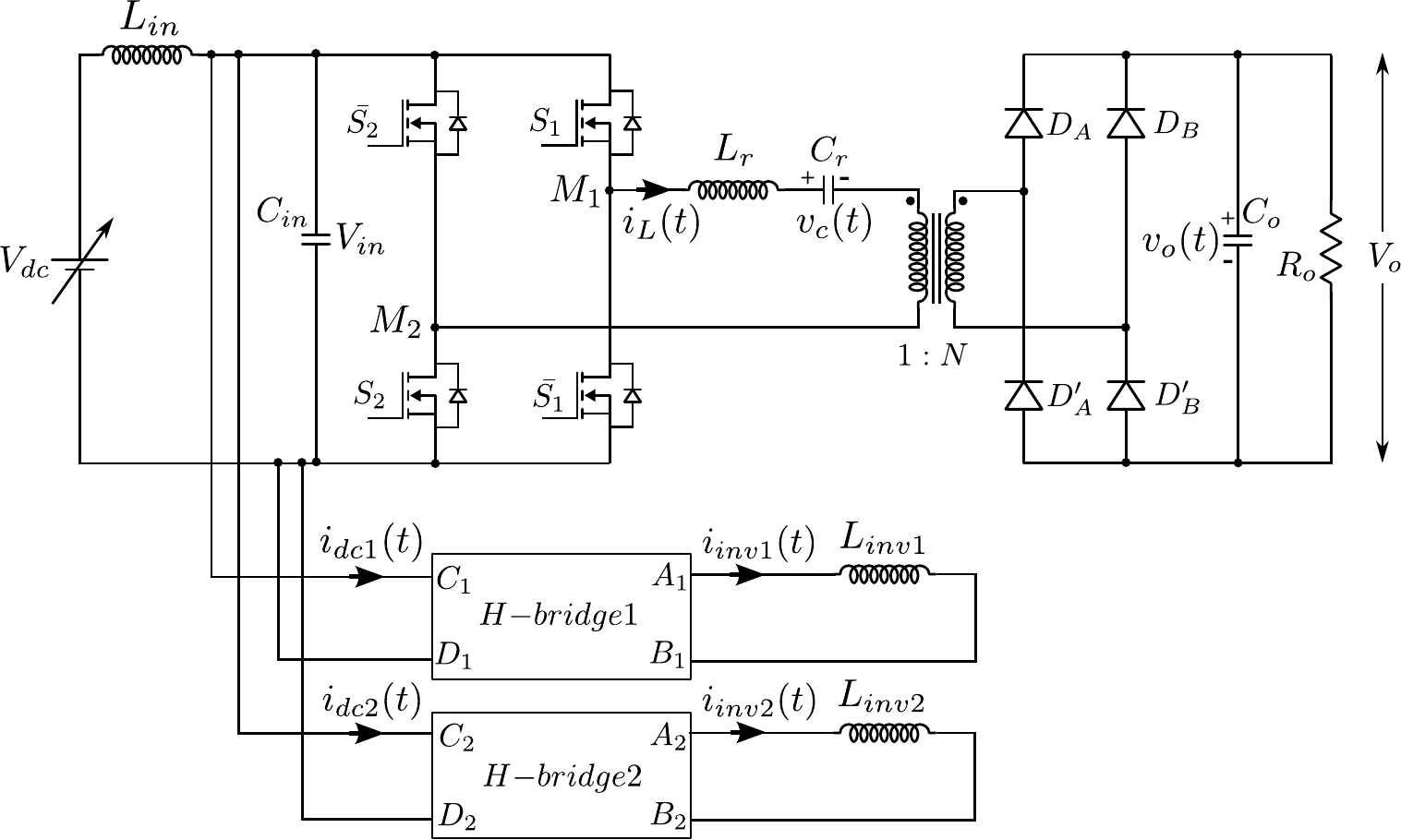}}\\
\subfloat[]{\includegraphics[keepaspectratio,scale=0.65]{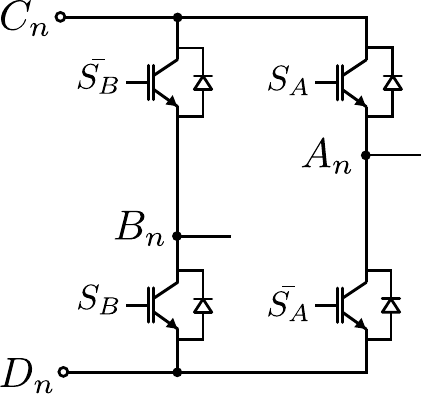}}\hspace{0.30in}
\subfloat[]{\includegraphics[keepaspectratio,scale=0.65]{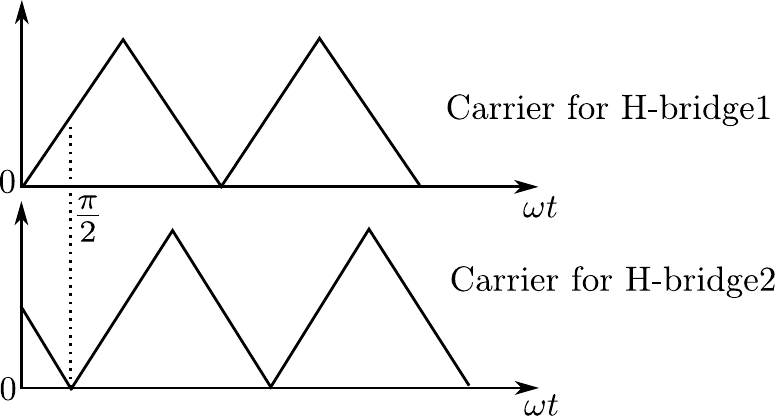}}
\caption{(a) Test circuit to evaluate the audiosusceptibility of SRC (b) schematic of H-bridge (c) carrier for H-bridge\textbf{\textit{1}} and H-bridge\textbf{\textit{2}}.}
\label{AStestckt}
\end{figure}

The test setup shown in Fig.~\ref{AStestckt}(a) is used to validate the audiosusceptibility response of the SRC. This setup consists of two H-bridge inverters and an SRC. The H-bridge is built using an IGBT, shown in Fig.~\ref{AStestckt}(b). The inductors $L_{inv1}$ and $L_{inv2}$ act as a load for H-bridge\textbf{\textit{1}} and H-bridge\textbf{\textit{2}} respectively. The objective of the H-bridge inverters are to introduce a known frequency of ripple in the dc voltage $V_{in}$ at the input of SRC. This is achieved by modulating the switches of the H-bridge with unipolar PWM, where the gate signals are generated by comparing the triangle carrier with sinusoidal modulating signal. The switching frequency of H-bridge is chosen as $10kHz$ and unipolar PWM pushes the effective switching frequency to $20kHz$. H-bridge\textbf{\textit{2}}, is also modulated as that of H-bridge\textbf{\textit{1}} except the carrier of H-bridge\textbf{\textit{2}} is shifted by $90^0$ from the carrier of H-bridge\textbf{\textit{1}} as shown in Fig.~\ref{AStestckt}(c). Such a phase shift pushes the effective switching frequency to $40kHz$, which is twice the individual H-bridge effective switching frequency.

\begin{table}[!t]
\renewcommand{\arraystretch}{1.3}
\caption{Values of components used in the audiosusceptibility test circuit}
\label{AStestcktvalues}
\centering
\begin{tabular}{|l|r|}
\hline
\bfseries Parameters & \bfseries Values\\
\hline
$L_{in}$ & $20mH$\\
\hline
$C_{in}$ & $900\mu F$\\
\hline
$L_{inv1}$, $L_{inv2}$ & $0.165mH$\\
\hline
Resonant inductor, $L_r$ & $150\mu H$\\
\hline
Resonant capacitor, $C_r$ & $16nF$\\
\hline
Output filter capacitor, $C_o$ & $100nF$\\
\hline
Load resistance, $R_o$ & $10k\Omega$\\
\hline
Turns ratio of transformer, $N$ & $16$\\
\hline
Leakage inductance of transformer referred to primary & $15\mu H$\\
\hline
Stray capacitance of transformer referred to primary, $C_s$ & $2.85nF$\\
\hline
\end{tabular}
\end{table}
For the audiosusceptibility test of SRC, the ripple frequency in $V_{in}$ is varied between $100Hz$ and $4kHz$. Hence the current drawn from the capacitor $C_{in}$ by H-bridge\textbf{\textit{1}} and H-bridge\textbf{\textit{2}}, referred to as $i_{dc1}$ and $i_{dc2}$ respectively, should be of frequency between $100Hz$ and $4kHz$. This is achieved by choosing the frequency of modulating signal between $50Hz$ and $2kHz$, which is also equal to the frequency of inverter current $i_{inv1}$ and $i_{inv2}$ shown in Fig.~\ref{AStestckt}(a). The higher value of effective switching frequency of $40kHz$ allows one to achieve a high ripple frequency of $4kHz$ with reduced unwanted frequencies. The inductor $L_{in}$ shown in Fig.~\ref{AStestckt}(a) connected in series with the variable dc source $V_{dc}$ ensures that ripple of frequency between $100Hz$ and $4kHz$ in $i_{dc1}$ and $i_{dc2}$ are primarily drawn only from the dc capacitor $C_{in}$. The corner frequency chosen for $L_{in}$ and $C_{in}$ is $37.5Hz$. The values of $L_{inv1}$ and $L_{inv2}$ are chosen to give an appreciable ripple in $V_{in}$. The SRC is designed for an output voltage of $10kV$dc for an input voltage of $625V$dc and nominal output power of $10kW$. The tank resonance frequency contributed by $L_r$ and $C_r$ is chosen as $102kHz$. The values of various components used in the H-bridge inverters and SRC are given in Table~\ref{AStestcktvalues}. Even though the rated input voltage of the SRC is $625V$dc, the audiosusceptibility test is carried out at a low input voltage around $10V$dc, considering that AS resonance conditions may be excited.

After turning on the SRC, the H-bridge\textbf{\textit{1}} and H-bridge\textbf{\textit{2}} are turned on to introduce a ripple in $V_{in}$. The frequency of the modulating signal is varied from $50Hz$ to $2kHz$, incremented in a step of $50Hz$. At each step, the ripple in input $V_{in}$ and output $V_o$ is recorded with an oscilloscope of $5Gs/s$ sampling rate. The built-in spectrum analysis function of the oscilloscope is used to find the magnitude and frequency spectrum of the sensed signal $V_{in}$ and $V_o$. By measuring the dc gain of SRC and from the measured magnitude of ripple in $V_{in}$ and $V_o$ the audiosusceptibility gain is computed by (\ref{gainASp}). The audiosusceptibility gain is plotted for each ripple frequency to find the SRC frequency response to input ripple.
\begin{figure}[!t]
\centering
\includegraphics[keepaspectratio,scale=0.35]{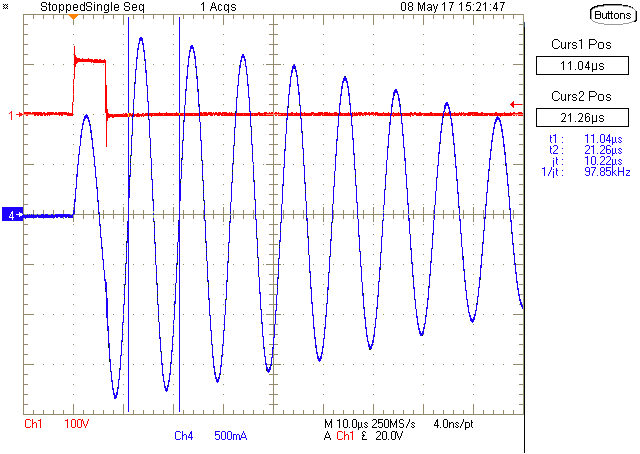}
\put(-140,143){\textsl{\footnotesize $i_{L}(t)$}}
\put(-215,143){\textsl{\footnotesize $v_{M_1M_2}(t)$}}
\caption{Experimental measurement with output load shows the impulse voltage $v_{M_1M_2}(t)$ and $i_{L}(t)$. SRC tank resonance frequency measured from $i_{L}(t)$ is also shown.} 
\label{actualfr}
\end{figure}
\subsection{Estimation of SRC tank resonance frequency ($f_r$)}
Even though the values of $L_r$ and $C_r$ are chosen for a tank resonance frequency of $102kHz$, due to the presence of magnetizing inductance and stray capacitance of the transformer the effective tank resonance frequency will be different from $102kHz$. An impulse signal is applied across $M_1$ and $M_2$ referred in Fig.~\ref{AStestckt}(a) to estimate the actual SRC tank resonance frequency. The experiment is conducted along with the dc side load $R_o$. The impulse signal is generated with the same test setup shown in Fig.~\ref{AStestckt}(a). The H-bridge inverters are kept in the off condition and the switches of SRC are triggered appropriately to generate an impulse voltage of $10\mu s$ duration across $M_1$ and $M_2$. The frequency of the resonant inductor current $i_L(t)$ is measured to compute the tank resonance frequency. Since the output capacitor $C_o$ charges in each resonance cycle, to avoid the influence of $v_o(t)$ on the estimation of tank resonance frequency, the frequency of the first full cycle is measured. Fig.~\ref{actualfr} shows the impulse waveform across $M_1$ and $M_2$, and the resonant inductor current measures a tank resonance frequency of $98kHz$. No significant change is observed in the value of tank resonance frequency, even when the experiment is repeated by shorting the secondary of the transformer. Assuming the influence of stray capacitance on the tank resonance frequency is negligible, $L_{reff}$ is computed from the measured tank resonance frequency by choosing $C_{reff}=C_r=16nF$, as given in Table~\ref{AStestcktvalues}. The $L_{reff}$ is computed as,
\begin{equation}
L_{reff}=\frac{1}{4\pi^2C_rf_r^2}=164.8\mu H
\end{equation}
Hence the SRC designed for the experiment will have characteristic impedance of,
\begin{equation}
Z_c=\sqrt{L_{reff}/C_r}=101.4\Omega
\end{equation}
Using the SRC parameters given in Table~\ref{AStestcktvalues}, the reflected resistance to the primary of the transformer having turns ratio $1:N$ is given by (\ref{rac})~\cite{text1}.
\begin{equation}\label{rac}
R_{ac}=\dfrac{8}{\pi^2}\dfrac{R_o}{N^2}=31.66\Omega
\end{equation}
Therefore, effective quality factor of SRC is given by,
\begin{equation}
Q_e=\dfrac{Z_c}{R_{ac}}=3.2
\end{equation}
For the experimental evaluation of audiosusceptibility, the $F$ value is chosen as $1.01$. Various derived characteristics of SRC are summarized in Table~\ref{SRCchar}.
\begin{table}[!t]
\renewcommand{\arraystretch}{1.3}
\caption{Various parameters of SRC that is used in the experiment}
\label{SRCchar}
\centering
\begin{tabular}{|l|r|}
\hline
\bfseries Parameters & \bfseries Values\\
\hline
Effective resonant inductor, $L_{reff}$ & $164.8\mu H$\\
\hline
Effective resonant capacitor, $C_{reff}$ & $16nF$\\
\hline
Characteristic impedance, $Z_c$ & $101.4\Omega$\\
\hline
Reflected resistance to the primary, $R_{ac}$ & $31.66\Omega$\\
\hline
Effective quality factor, $Q_e$ & $3.2$\\
\hline
Resonance frequency, $f_r$ & $98kHz$\\
\hline
Ratio of switching to SRC tank resonance frequency, $F$ & $1.01$\\
\hline
\end{tabular}
\end{table}
\subsection{Audiosusceptibility experimental results}
\begin{figure}[!t]
\centering
\includegraphics[keepaspectratio,scale=0.33]{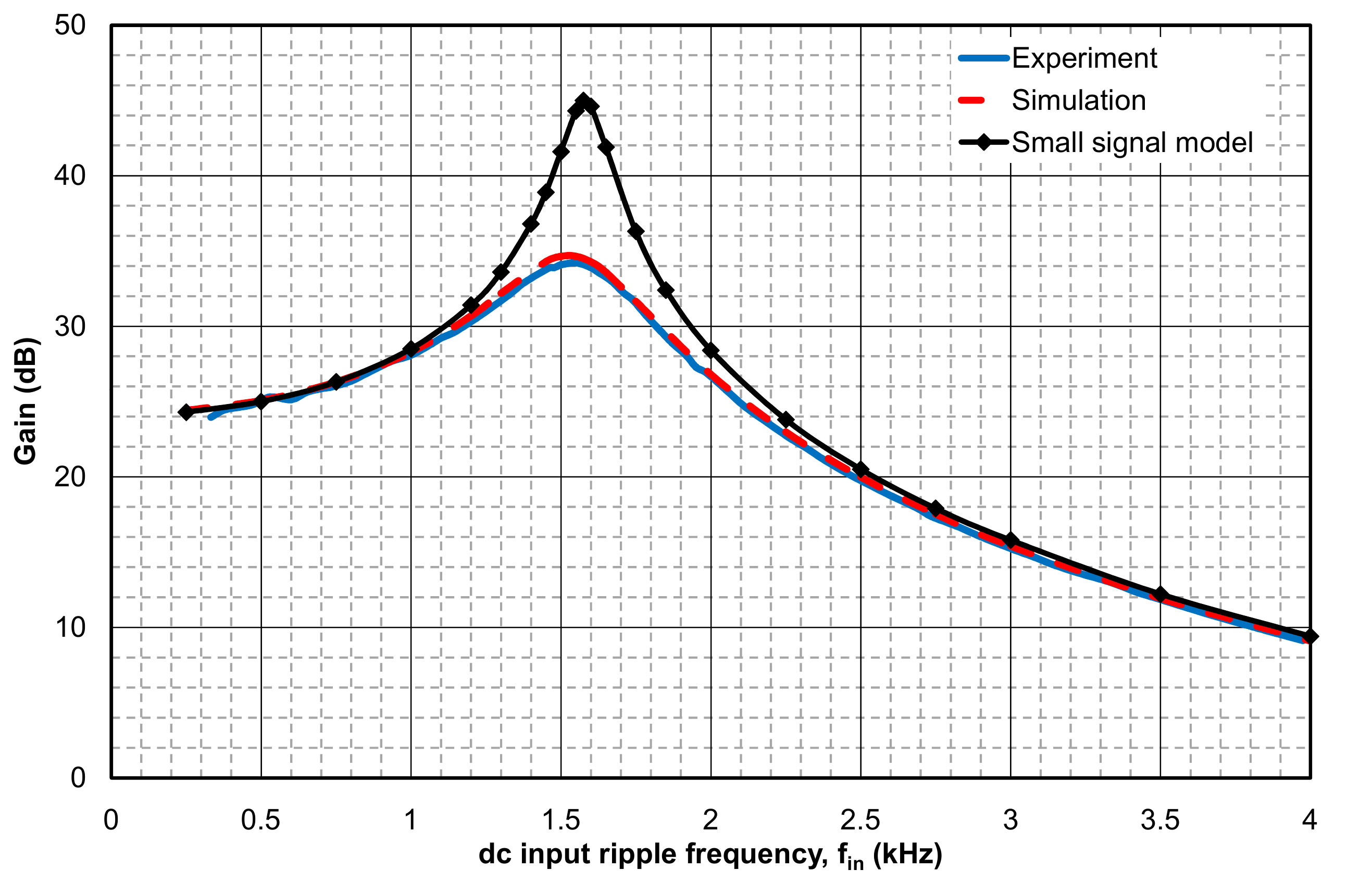}
\caption{Comparison of audiosusceptibility gain from experiment, simulation and small signal model of SRC.} 
\label{asgaincmp}
\end{figure}

The input dc voltage applied for the experiment is $8.4V$ results in an output voltage of $133.5V$. This gives a dc gain of $15.9$. The input ripple frequency in $V_{in}$ is varied from $100Hz$ to $4kHz$. From the measured magnitude of input and output ripple and the obtained dc gain, by (\ref{gainASp}) the audiosusceptibility gain is computed at each frequency. Fig.~\ref{asgaincmp} shows the experimentally obtained audiosusceptibility gain with the dc input ripple frequency. AS resonance is observed in Fig.~\ref{asgaincmp} at frequency $1550Hz$. Substituting the parameters given in Table~\ref{SRCchar} into (\ref{winr}), the AS resonance frequency for input ripple is obtained analytically as $1570Hz$.

\begin{figure}[!t]
\centering
\includegraphics[keepaspectratio,scale=0.39]{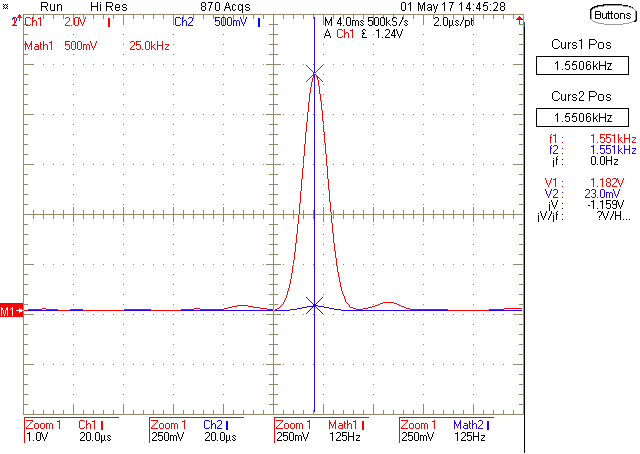}
\put(-120,135){\textsl{\footnotesize Output\textit{~(250mV/div.)}}}
\put(-120,30){\textsl{\footnotesize Input\textit{~(250mV/div.)}}}
\put(-105,40){\vector(-1,1){18}}
\put(-197,143){\vector(1,0){15}}
\put(-187,143){\vector(-1,0){15}}
\put(-202,137){\textsl{\tiny $125Hz$}}
\caption{Magnitude spectrum of input (Math2) and output (Math1) ripple corresponds to a frequency of $1551Hz$.} 
\label{magspectrum1}
\end{figure}

\begin{figure}[!t]
\centering
\includegraphics[keepaspectratio,scale=0.39]{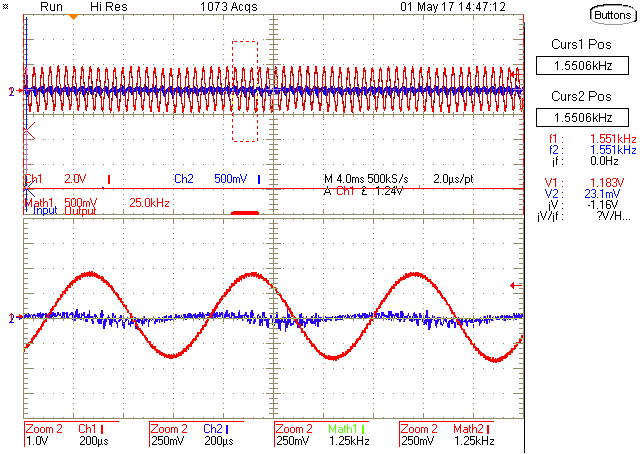}
\put(-115,75){\textsl{\footnotesize Output\textit{~(1V/div.)}}}
\put(-120,25){\textsl{\footnotesize Input\textit{~(250mV/div.)}}}
\put(-105,32){\vector(1,1){18}}
\caption{AC coupled input (CH2) and output (CH1) ripple in time domain corresponds to a frequency of $1551Hz$.}
\label{rippletime}
\end{figure}

The magnitude spectrum of the input ripple and output ripple obtained from experiment at input ripple frequency of $1550Hz$ which corresponds to the AS resonance frequency is shown in Fig.~\ref{magspectrum1}. The rms values of output and input ripple measured are $1.182V$ and $23mV$ respectively, gives a gain of $34.2dB$. The time domain input and output ripple recorded from the experiment for the frequency of $1550Hz$ is shown in Fig.~\ref{rippletime}. Fig.~\ref{asgaincmp} also shows that beyond $2114Hz$ the gain offered by SRC to input ripple is less than unity as the gain at $2114Hz$ is $23dB$ which is equal to the turns ratio of the transformer.

Various SRC component values given in Table~\ref{AStestcktvalues} are used in the simulation to find the  audiosusceptibility gain for various values of input ripple frequency. Similarly the audiosusceptibility gain is computed with the small signal audiosusceptibility model derived in (\ref{TF}). The audiosusceptibility gain plot obtained from experiment, simulation and small signal model are compared in Fig.~\ref{asgaincmp}. The audiosusceptibility gain at resonance condition obtained from experiment, simulation and small signal model are $34.2dB$, $34.6dB$ and $44dB$ respectively. The frequency at which maximum audiosusceptibility gain occurred in experiment, simulation and small signal model is found to be closely matching and the values are $1550Hz$, $1550Hz$ and $1570Hz$ respectively. From Fig.~\ref{asgaincmp} the gain contributed by the small signal model is higher than that obtained from experiment and simulation. The audiosusceptibility small signal model is derived without considering the circuit elements such as the stray capacitance of the transformer, $C_s$, winding resistance and transformer magnetizing inductance.

\begin{figure}[!t]
\centering
\includegraphics[keepaspectratio,scale=0.33]{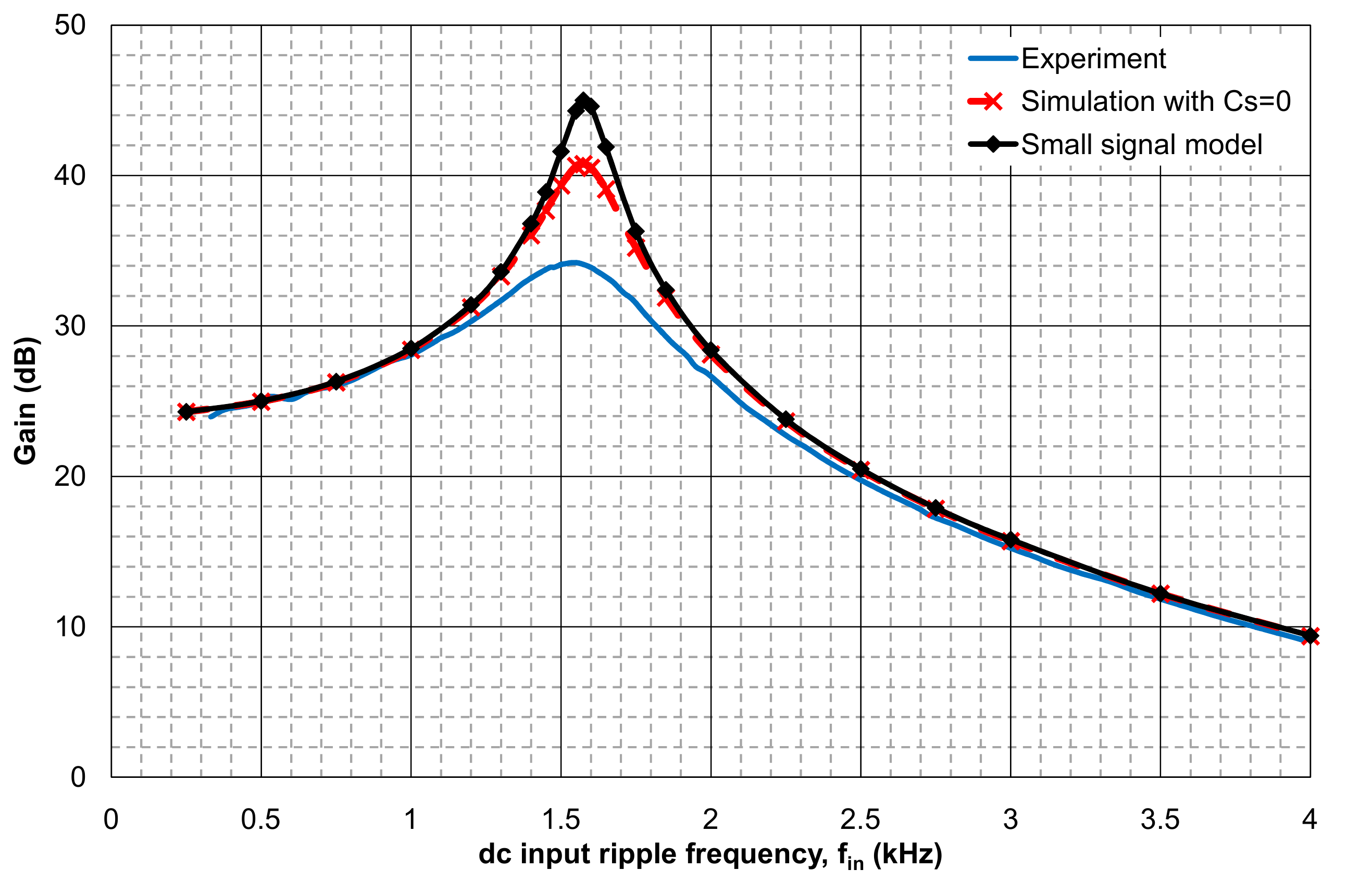}
\caption{Comparison of audiosusceptibility gain from experiment, simulation with $C_s=0$ and small signal model of SRC.} 
\label{asgaincmpCso}
\end{figure}
\begin{table}[!h]
\renewcommand{\arraystretch}{1.3}
\caption{Audiosusceptibility resonance frequency and the gain at resonance condition obtained from experiment, simulation with and without $C_s$ and small signal model}
\label{gainfripple}
\centering
\begin{tabular}{|l|r|r|r|r|}
\hline
\bfseries Parameters & \bfseries Exper-& \multicolumn{2}{c|}{\bfseries Simulation}& \multicolumn{1}{c|}{\bfseries AS Small} \\
\cline{3-4}
& \bfseries iment& \bfseries with $C_s$&\bfseries with $C_s{=}0$&\bfseries signal model \\
\hline
AS gain at  & $34.2$&$34.6$&$41.0$&$44.0$\\
resonance (dB)&&&&\\
\hline
AS resonance & $1550$&$1550$&$1575$&$1570$\\
frequency (Hz)&&&&\\
\hline
\end{tabular}
\end{table}

To verify this condition, the simulations are repeated by keeping the stray capacitance of the transformer to zero value. The audiosusceptibility gain obtained by experiment, simulation with $C_s{=}0$ and small signal model are compared in Fig.~\ref{asgaincmpCso}. Fig.~\ref{asgaincmpCso} shows that the audiosusceptibility gain with simulation, keeping the stray capacitance of the transformer to zero, is close to the audiosusceptibility gain plot from the small signal model. Values for audiosusceptibility resonance frequency and the gain at resonance condition obtained from experiment, simulation with and without $C_s$ and small signal model are tabulated in Table~\ref{gainfripple}. The gain obtained from small signal model is close to gain from simulation with $C_s{=}0$. Also by introducing $C_s$ in the simulation, the gain obtained from simulation is close to the experimental results. This shows that the small signal model is valid considering that it is formulated after neglecting parasitic elements. All results from the experiment are closely matching with the analytical results derived in section~\ref{sec3}.
\section{Conclusion}
The paper formulates a method to handle the $3$ state variables of an SRC while obtaining state space model using exact discretization modelling approach. From the non-linear state space model a small signal state space model is derived. Considering practically observed assumptions, the small signal model is simplified and the audiosusceptibility transfer function is derived. The analysis of AS transfer function shown a possibility of resonance peak and an expression relating the AS resonance frequency for input ripple with different SRC components is derived. The frequency response of AS transfer function is verified by simulation and compared with the derived analytical expressions for different values of $Q_e$ and $F$. From further analysis a region is identified that connects $Q_e$ and $F$ with the ``less than unity gain'' audiosusceptibility gain. The obtained region is verified by time domain simulation results and such region helps in the design of SRC converters with superior audiosusceptibility performance. A test setup for the experimental evaluation of AS transfer function is also proposed in the paper. The influence of stray parameters on AS characteristics is discussed. All the analysis on analytical model, simulation and experimental results on a $10kW$ SRC are found to be closely matching.

\begin{strip}
\appendix
The elements of the matrices $A_d$ and $B_d$ are given in (\ref{Ad}) and (\ref{Bd}) respectively. The elements of $T_d$ is given in (\ref{td}) where as elements of  $T_{kx}$ and $T_{ku}$ are given in (\ref{tk}).
\renewcommand{\theequation}{A.\arabic{equation}}
\setcounter{equation}{0}  
\begin{align}
\hline\nonumber\\
&\mathbf{a_{11}}{=}\cos\omega_rT_s,\enskip \mathbf{a_{12}}{=}{-}\dfrac{\sin\omega_rT_s}{Z_c},\enskip \mathbf{a_{21}}{=}Z_c\sin\omega_rT_s,\enskip \mathbf{a_{22}}{=}\cos\omega_rT_s,\enskip \mathbf{a_{13}}{=}\dfrac{\sin\omega_rT_s{+}2\sin\omega_r(T_s{-}T_3){-}2\sin\omega_r (T_s{-}T_1)}{NZ_c},\nonumber\\
&\mathbf{a_{23}}{=}\dfrac{1{-}\cos\omega_rT_s{-}2\cos\omega_r(T_s{-}T_3){+}2\cos\omega_r(T_s{-}T_1)}{N},\enskip\mathbf{b_{11}}{=}\dfrac{\sin\omega_rT_s-2\sin\omega_r\dfrac{T_s}{2}}{Z_c},\enskip\mathbf{b_{21}}{=}2\cos\omega_r\dfrac{T_s}{2}-\cos\omega_rT_s-1\nonumber\\ &\text{where}~~\omega_r=1/\sqrt{L_rC_r}, Z_c=\sqrt{L_r/C_r}\label{Ad}\nonumber\\
\\
\hline\nonumber\\
&\mathbf{a_{31}}{=}G_1\left[2e^{\tfrac{T_1}{R_oC_o}}g_1^{\prime}(T_1)-2e^{\tfrac{T_3}{R_oC_o}}g_1^{\prime}(T_3)+e^{\tfrac{T_s}{R_oC_o}}g_1^{\prime}(T_s)-\dfrac{1}{R_oC_o}\right]\nonumber\\
&\mathbf{a_{32}}{=}\dfrac{G_1}{Z_c}\left[2e^{\tfrac{T_3}{R_oC_o}}g_1(T_3)-2e^{\tfrac{T_1}{R_oC_o}}g_1(T_1)-e^{\tfrac{T_s}{R_oC_o}}g_1(T_s)-\omega_r\right]\nonumber\\
&\mathbf{a_{33}}{=}-e^{\tfrac{T_s}{R_oC_o}}+\dfrac{G_1}{NZ_c}\biggl[ 2e^{\tfrac{T_1}{R_oC_o}}g_1(T_1)-2e^{\tfrac{T_3}{R_oC_o}}g_1(T_3)+e^{\tfrac{T_s}{R_oC_o}}g_1(T_s)+\omega_r+4e^{\tfrac{T_3}{R_oC_o}}g_1(T_3-T_1)\nonumber\\
&-2e^{\tfrac{T_s}{R_oC_o}}g_1(T_s-T_1)+2\omega_re^{\tfrac{T_1}{R_oC_o}}+2e^{\tfrac{T_s}{R_oC_o}}g_1(T_s-T_3)+2\omega_re^{\tfrac{T_3}{R_oC_o}}\biggr]\nonumber\\
&\mathbf{b_{31}}{=}\dfrac{G_1}{Z_c}\biggl[ 2e^{\tfrac{T_1}{R_oC_o}}g_1(T_1){-}2e^{\tfrac{T_3}{R_oC_o}}g_1(T_3){+}e^{\tfrac{T_s}{R_oC_o}}g_1(T_s)+\omega_r+4e^{\tfrac{T_3}{R_oC_o}}g_1(T_3-\tfrac{T_s}{2})-2e^{\tfrac{T_s}{R_oC_o}}g_1(\tfrac{T_s}{2})+2\omega_re^{\tfrac{T_s/2}{R_oC_o}}\biggr]\nonumber\\
&\text{where}~~g_1(\boldsymbol{\cdot})=\dfrac{1}{R_oC_o}\sin\omega_r(\boldsymbol{\cdot})-\omega_r\cos\omega_r(\boldsymbol{\cdot}),~~ g_1^{\prime}(\boldsymbol{\cdot})=\dfrac{dg_1(\boldsymbol{\cdot})}{d\omega_r(\boldsymbol{\cdot})},~~G_1=-\dfrac{1}{NC_o}e^{-\tfrac{T_s}{R_oC_o}}\dfrac{1}{\left(\tfrac{1}{R_oC_o}\right)^2+\omega_r^2}\label{Bd}\nonumber\\
\\
\hline\nonumber\\
&\mathbf{td_{11}}{=}\dfrac{2\omega_rV_o\cos\omega_r(T_s{-}T_1)}{NZ_c}{,}\enskip\mathbf{td_{12}}{=}{-}\dfrac{2\omega_rV_o\cos\omega_r(T_s{-}T_3)}{NZ_c}{,}\enskip\mathbf{td_{21}}{=}\dfrac{2\omega_rV_o\sin\omega_r(T_s{-}T_1)}{N}{,}\mathbf{td_{22}}{=}{-}\dfrac{2\omega_rV_o\sin\omega_r(T_s{-}T_3)}{N}\nonumber\\
&\mathbf{td_{31}}{=}-\dfrac{2I_L\cos\omega_rT_1}{NC_o}e^{\tfrac{-(T_s-T_1)}{R_oC_o}}+\dfrac{2V_c\sin\omega_rT_1}{NZ_cC_o}e^{\tfrac{-(T_s-T_1)}{R_oC_o}}+\dfrac{GV_o}{NZ_c}\biggl[ 2G_2e^{\tfrac{T_1}{R_oC_o}}\sin\omega_rT_1-4\omega_re^{\tfrac{T_3}{R_oC_o}}g_1^{\prime}(T_3-T_1)+\nonumber\\
&\qquad\quad 2\omega_re^{\tfrac{T_s}{R_oC_o}}g_1^{\prime}(T_s-T_1)+\dfrac{2\omega_r}{R_oC_o}e^{\tfrac{T_1}{R_oC_o}}\biggr]-\dfrac{2V_{in}\sin\omega_rT_1}{NZ_cC_o}e^{\tfrac{-(T_s-T_1)}{R_oC_o}}\nonumber\\
&\mathbf{td_{32}}{=}\dfrac{2I_L\cos\omega_rT_3}{NC_o}e^{\tfrac{{-}(T_s{-}T_3)}{R_oC_o}}{-}\dfrac{2V_c\sin\omega_rT_3}{NZ_cC_o}e^{\tfrac{{-}(T_s{-}T_3)}{R_oC_o}}{+}\dfrac{GV_oe^{\tfrac{T_3}{R_oC_o}}}{NZ_c}\biggl[ {-}2G_2\sin\omega_rT_3{+}\dfrac{2\omega_r}{R_oC_o}{+}\nonumber\\
&\qquad\quad 4G_2\sin\omega_r(T_3{-}T_1){-}2\omega_re^{\tfrac{(T_s{-}T_3)}{R_oC_o}}g_1^{\prime}(T_s{-}T_3)\biggr]{+}\dfrac{GV_{in}e^{\tfrac{T_3}{R_oC_o}}}{Z_c}\biggl[ {-}2G_2\sin\omega_rT_3{+}\dfrac{4g_2(T_3{-}\tfrac{T_s}{2})}{R_oC_o}{-}4\omega_rg_2^{\prime}(T_3{-}\tfrac{T_s}{2})\biggr]\nonumber\\
&\text{where}~~g_2(\boldsymbol{\cdot})=\dfrac{1}{R_oC_o}\sin\omega_r(\boldsymbol{\cdot})+\omega_r\cos\omega_r(\boldsymbol{\cdot}),~~g_2^{\prime}(\boldsymbol{\cdot})=\dfrac{dg_2(\boldsymbol{\cdot})}{d\omega_r(\boldsymbol{\cdot})},~~G_2=\left(\tfrac{1}{R_oC_o}\right)^2+\omega_r^2\label{td}\nonumber\\
\\
\hline\nonumber\\
&\mathbf{tx_{11}}{=}\dfrac{\cos\omega_rT_1}{f^{\prime}_{T1}},\quad\mathbf{tx_{12}}{=}-\dfrac{\sin\omega_rT_1}{Z_cf^{\prime}_{T1}},\quad\mathbf{tx_{13}}{=}\dfrac{\sin\omega_rT_1}{NZ_cf^{\prime}_{T1}},\quad\mathbf{tx_{21}}{=}\dfrac{\cos\omega_rT_3}{f^{\prime}_{T3}}+\dfrac{2V_o\omega_r\cos\omega_r(T_3-T_1)\cos\omega_rT_1}{NZ_cf^{\prime}_{T1}f^{\prime}_{T3}}\nonumber\\
&\mathbf{tx_{22}}{=}{-}\dfrac{\sin\omega_rT_3}{Z_cf^{\prime}_{T3}}{-}\dfrac{2V_o\omega_r\cos\omega_r(T_3{-}T_1)\sin\omega_rT_1}{NZ_c^2f^{\prime}_{T1}f^{\prime}_{T3}}{,}\quad\mathbf{tx_{23}}{=}\dfrac{\sin\omega_rT_3{-}2\sin\omega_r(T_3{-}T_1)}{NZ_cf^{\prime}_{T3}}{+}\dfrac{2V_o\omega_r\cos\omega_r(T_3{-}T_1)\sin\omega_rT_1}{N^2Z_c^2f^{\prime}_{T1}f^{\prime}_{T3}}\nonumber\\
&\mathbf{tu_{11}}{=}\dfrac{\sin\omega_rT_1}{Z_cf^{\prime}_{T1}},\mathbf{tu_{21}}{=}\dfrac{\sin\omega_rT_3{-}2\sin\omega_r(T_3{-}\tfrac{T_s}{2})}{Z_cf^{\prime}_{T3}}{+}\dfrac{2V_o\omega_r\cos\omega_r(T_3-T_1)\sin\omega_rT_1}{NZ_c^2f^{\prime}_{T1}f^{\prime}_{T3}}\nonumber\\
&\text{where}~f_{T1}^{\prime}{=}-\dfrac{f_{T1}}{dT_1},~f_{T3}^{\prime}{=}-\dfrac{f_{T3}}{dT_3}\label{tk}\nonumber\\
\\
\hline\nonumber
\end{align}
\end{strip}
\section*{Acknowledgment}
This work is supported by Ministry of Electronics and Information Technology, Govt. of India, through NaMPET programme.
%
%
%
%
%


%
%
%

\bibliographystyle{IEEEtran}
\bibliography{Reference}

\end{document}